\documentclass[twocolumn]{aastex631}

\graphicspath{{./}{figures/}}
\usepackage{lipsum}
\usepackage{graphicx}
\usepackage{subfigure}
\usepackage{comment}
\usepackage{booktabs}
\usepackage{amsmath}
\usepackage{etoolbox}
\usepackage{IEEEtrantools}
\usepackage{parskip}
\usepackage{ragged2e}
\usepackage{textcomp,gensymb}

\begin{document}
\UseRawInputEncoding
\title{NEW METRICS TO PROBE THE DYNAMICAL STATE OF GALAXY CLUSTERS}

\correspondingauthor{Nico Cappelluti}
\email{ncappelluti@miami.edu}

\author{Giulia Cerini}
\affiliation{Department of Physics, University of Miami, 1320 S Dixie Hway, Coral Gables, FL 33146, USA}

\author{Nico Cappelluti} 
\affiliation{Department of Physics, University of Miami, Coral Gables, FL 33124, USA}

\author{Priyamvada Natarajan}
\affiliation{Department of Astronomy, Yale University, 52 Hillhouse Avenue, New Haven, CT 06511, USA}
\affiliation{Department of Physics, Yale University, P.O. Box 208121, New Haven, CT 06520, USA}
\affiliation{Black Hole Initiative, Harvard University, 20 Garden Street, Cambridge, MA 02138, USA}

\begin{abstract} \label{abstract}
We present new diagnostic metrics to probe the dynamical state of galaxy clusters. These novel metrics rely on the computation of the power spectra of the matter and gas distributions and their cross-correlation derived from cluster observations. This analysis permits us to cross-correlate the fluctuations in the matter distribution, inferred from high-resolution lensing mass maps derived from Hubble Space Telescope (HST) data, with those derived from the emitted X-ray surface brightness distribution of the hot Intra-Cluster Medium (ICM) from the Chandra X-ray Observatory (CXO). These methodological tools allow us to quantify with unprecedented resolution the coherence with which the gas traces the mass and interrogate the assumption that the gas is in hydro-static equilibrium with the underlying gravitational potential. We characterize departures from equilibrium as a function of scale with a new gas-mass coherence parameter. The efficacy of these metrics is demonstrated by applying them to the analysis of two representative clusters known to be in different dynamical states: the massive merging cluster Abell 2744, from the HST Frontier Fields (HSTFF) sample, and the dynamically relaxed cluster Abell 383, from the Cluster Lensing And Supernova Survey with Hubble (CLASH) sample. Using lensing mass maps in combination with archival Chandra data, and simulated cluster analogs available from the OMEGA500 suite, we quantify the fluctuations in the mass and X-ray surface brightness and show that new insights into the dynamical state of the clusters can be obtained from our gas-mass coherence analysis. 
\end{abstract}


\section{Introduction} \label{sec:intro}

Clusters of galaxies are the most massive and recently assembled structures in the Universe and are the largest repositories of collapsed dark matter in the Universe, with masses in the range of $~10^{14-15}M_{\odot}$. Although the gravitational potential of clusters is dominated by the non-baryonic dark matter component, they contain X-ray emitting hot gas with typical temperatures of $10^{7-8}K$, the ICM. In the smooth and static gravitational potential provided by the dark matter, the ICM gas is expected to be in hydro-static equilibrium, well aligned with the underlying equipotential surfaces, with a smooth density and temperature distribution. Under such conditions, the detected emission of X-ray surface brightness is interpreted as reflecting the fidelity of the gas in tracing the gravitational potential. However, this assumption of hydro-static equilibrium is unlikely to hold at all times, as departures from equilibrium are expected in assembling clusters during on-going merging activity. Besides, other sources of non-thermal pressure support for the gas, like those provided by magnetic fields, plasma instabilities, shocks, motions of galaxies and Active Galactic Nuclei (AGN) feedback, are also expected to contribute. These dynamical departures from equilibrium are expected to leave spatial imprints that can be quantified. For instance, on the scale of $\sim$ 100 kpc, the gas is mainly perturbed by mergers, while plasma instabilities are expected to produce disturbances on much smaller sub-kpc scales. On intermediate scales the gas is stirred and disturbed by the motions of galaxies as well as energy injected by the accreting central black holes.

Several previous studies have explored these perturbations and their impact on a wide range of spatial scales, and have used them in turn to study and quantify the physical processes that operate in the ICM (e.g. \citealt{Churazov_2012}, \citealt{zhuravleva17}). These studies, as well as other investigations of galaxy cluster dynamics, are still always reliant on two underlying assumptions: (i) clusters are in hydro-static equilibrium and (ii) their multiple components, namely the gas and member galaxies, are unbiased tracers of the underlying gravitational potential. Multiple studies of on-going merging clusters have revealed that these assumptions are likely invalid (e.g. \citealt{Markevitch_2002}, \cite{Clowe_2006}, \citealt{Emery_2017}). For instance, in the extreme case of the Bullet cluster it is evident that the collisional gas does not trace the collisionless dark matter (\citealt{Clowe_2006}), as the dark matter clumps from two merging sub-clusters are spatially well separated from the dissipative hot ICM. 

While the lensing effects of a galaxy cluster are independent of its dynamical state, it is found that the most efficient observed lenses are actively assembling and are therefore often dynamically complex and likely to be out of equilibrium (\citealt{Lotz_2017}). The most efficient cluster lenses tend to be massive merging clusters replete with in-falling substructures, and out of hydro-static equilibrium. While this has been known for more than a decade, recent work from detailed combined X-ray and lensing analysis of several HSTFF clusters (see for instance work by \citealt{Jauzac_2018}) has clearly demonstrated that massive cluster-lenses tend to contain abundant substructures and are neither relaxed nor in hydro-static equilibrium. 

Quantifying mass distributions from lensing has permitted mapping the detailed spatial distribution of matter and has recently revealed tensions with expectations and predictions from the cold dark matter paradigm. In a recent study of HST cluster lenses by \citealt{Meneghetti_2020}, they report a discrepancy between the observed efficiency of strong lensing on small scales by cluster member galaxies and predictions from simulations of clusters in the standard cold dark matter model. Careful and detailed studies of small-scale power are therefore warranted. The new methodology we develop here offers yet another metric to quantify substructures in the mass and in the gas and compare with cosmological simulations.

In this paper, we present the methods and demonstrate the proof of concept with results of the application of a new, fast and efficient method to assess the dynamical state of galaxy clusters. In the power spectrum and gas-mass coherence analysis developed here, we cross-correlate the fluctuations in the matter and X-ray surface brightness distributions, and use this to quantify how well the gas traces the underlying dark matter potential. Before extending our analysis to a wider sample of galaxy clusters in future work, in order to assess the validity and feasibility of applying this new method, we analyze two galaxy clusters, that bracket the wide range of dynamical states, as demonstrated in previous studies: the massive merging HSTFF cluster Abell 2744 (that is clearly out of equilibrium), and the cool core relaxed CLASH cluster Abell 383 (that is expected to be in equilibrium). The X-ray maps for both clusters are obtained from the publicly available Chandra archival data. Throughout the paper errors are quoted at 1$\sigma$ level and fluxes refer to the 0.5-7.0 keV band. In addition, our observationally derived power spectra are compared with those obtained from simulated mass analogs in the OMEGA500\footnote{\RaggedRight\url{http://gcmc.hub.yt/omega500/}} cosmological suite (see Table \ref{table:1}). It is worth highlighting that power spectrum analysis has become the tool of choice in investigations in X-ray and other wavelengths (e.g. \citealt{Kashlinsky2005}, \citealt{Hickox_2006}, \citealt{Cappelluti_2012}, \citealt{Churazov_2012}, \citealt{Kashlinsky_2012}, \citealt{Cappelluti_2013}, \citealt{Helgason_2014}, \citealt{Cappelluti_2017}, \citealt{Eckert_2017}, \citealt{zhuravleva17}, \citealt{Li_2018}, \citealt{Kashlinsky_2018}), as it allows an easy disentangling of the various spatial components of a signal, and the determination of their origin. Furthermore, the standard Fourier analysis allows us to spatially resolve fluctuations and study their coherence as a function of scale with a resolution not easily achievable with other methods. The application of these analysis techniques, as we show, can reveal the presence of unrelaxed regions in clusters that were believed to be in hydro-static equilibrium, as we report in the following sections.

The paper is structured as follows: the data are described in Section \ref{data}. We briefly summarize the mass reconstruction of Abell 2744 and Abell 383 in Subsection \ref{lensing}, the X-ray data reduction in Subsection \ref{xray}, and the sample of OMEGA500 simulations in Subsection \ref{omega500}. We present our methodology and derive the new metrics in Section \ref{formalism}, show the results of our analysis in Sections \ref{results} and close with a discussion of conclusions and future prospects in the Section \ref{conclusions}.

 \begin{table*}
\centering
\begin{tabular}{@{}ccccc@{}}
    \toprule
 Target & Virial mass ($M_{SUN}$) & Virial radius ($Mpc$) & Redshift & Label\\
    \midrule
 ABELL2744 & $3.10\times 10^{15}$ & $2.80 $ & $0.308$ & U  \\
 ABELL383 & $7.67\times 10^{14}$ & $2.10$ & $0.187$ & R\\
 HALO C-1 & $6.10\times 10^{14}$ & $1.76$ & $0.000$ & U\\
 HALO C-2 & $5.68\times 10^{14}$ & $1.71$ & $0.000$ & U\\
 HALO C-3 & $6.87\times 10^{14}$ & $1.83$ & $0.000$ & R\\
 HALO C-4 & $1.60\times 10^{15}$ & $2.42$ & $0.000$ & U\\
     \bottomrule
\end{tabular}
\caption{Sample of observed and simulated targets used in this investigation. The table entries are: cluster name (the first column), virial mass (second column), virial radius (third column), redshift (fourth column), and dynamical/morphological state classified as relaxed, denoted by R, and unrelaxed, denoted by U, respectively (fifth column). }
\label{table:1}
\end{table*}

\begin{figure*}
    \begin{center}
            \subfigure[]{\includegraphics[width=0.35\textwidth]{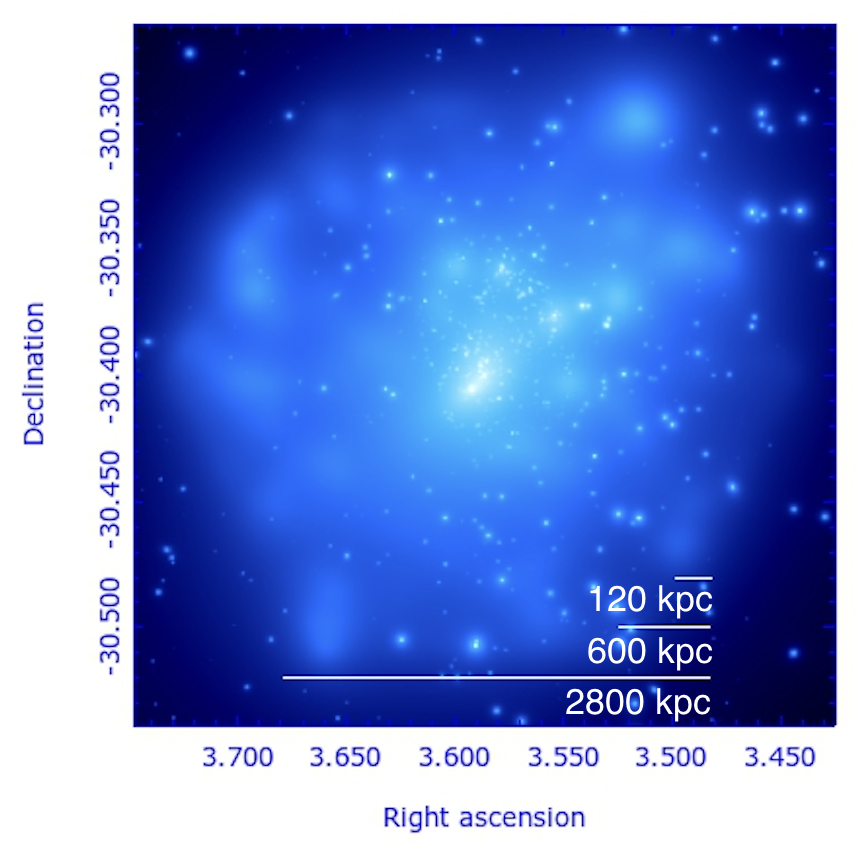}}
            \subfigure[]{\includegraphics[width=0.35\textwidth]{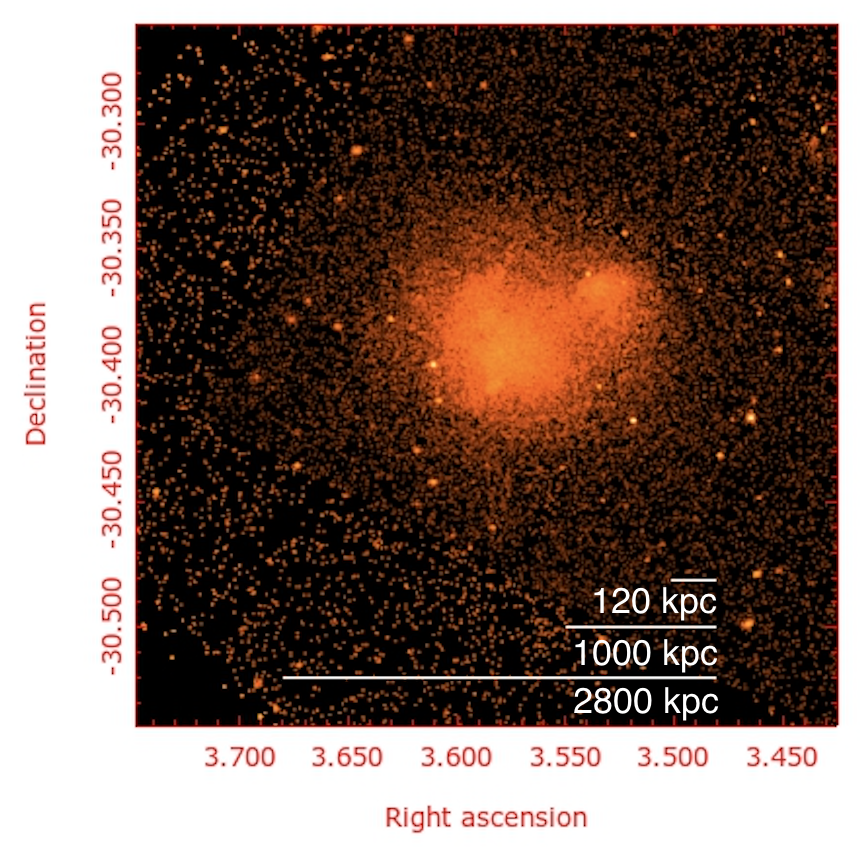}}  
            \subfigure[]{\includegraphics[width=0.3\textwidth]{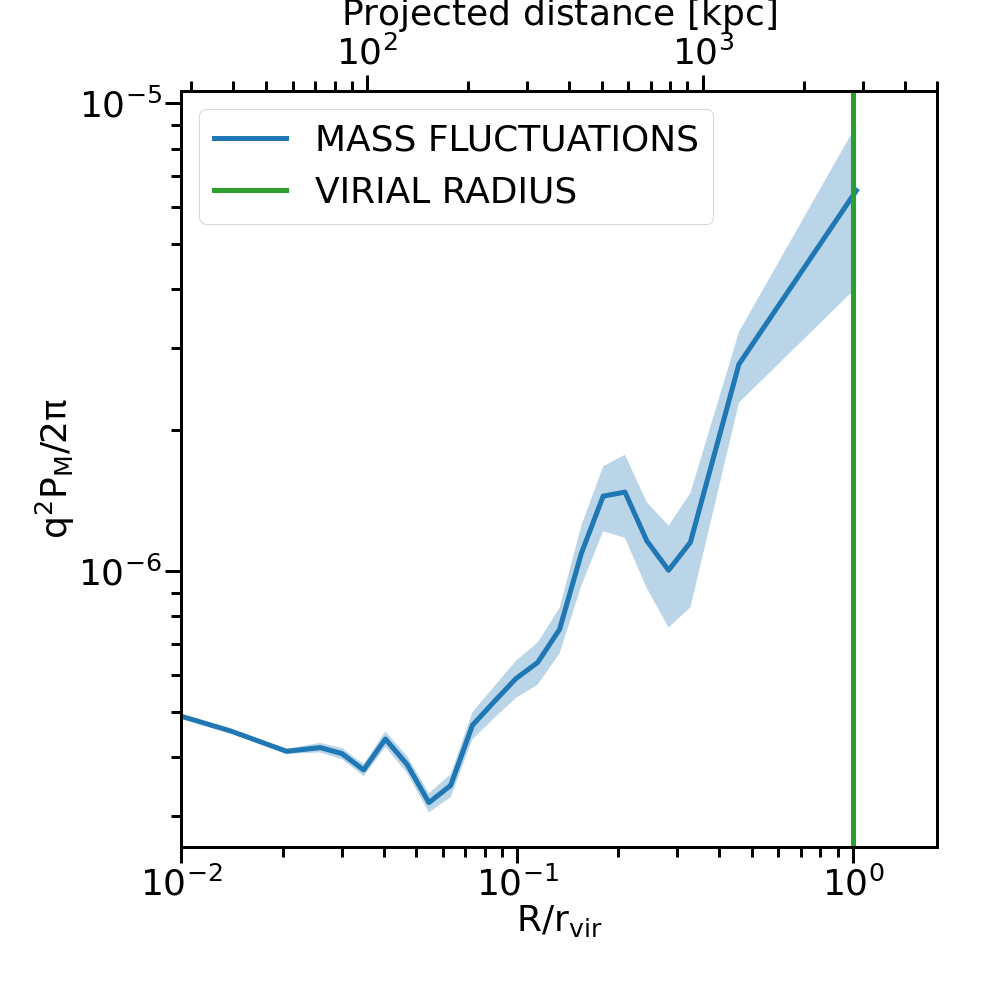}} 
            \subfigure[]{\includegraphics[width=0.3\textwidth]{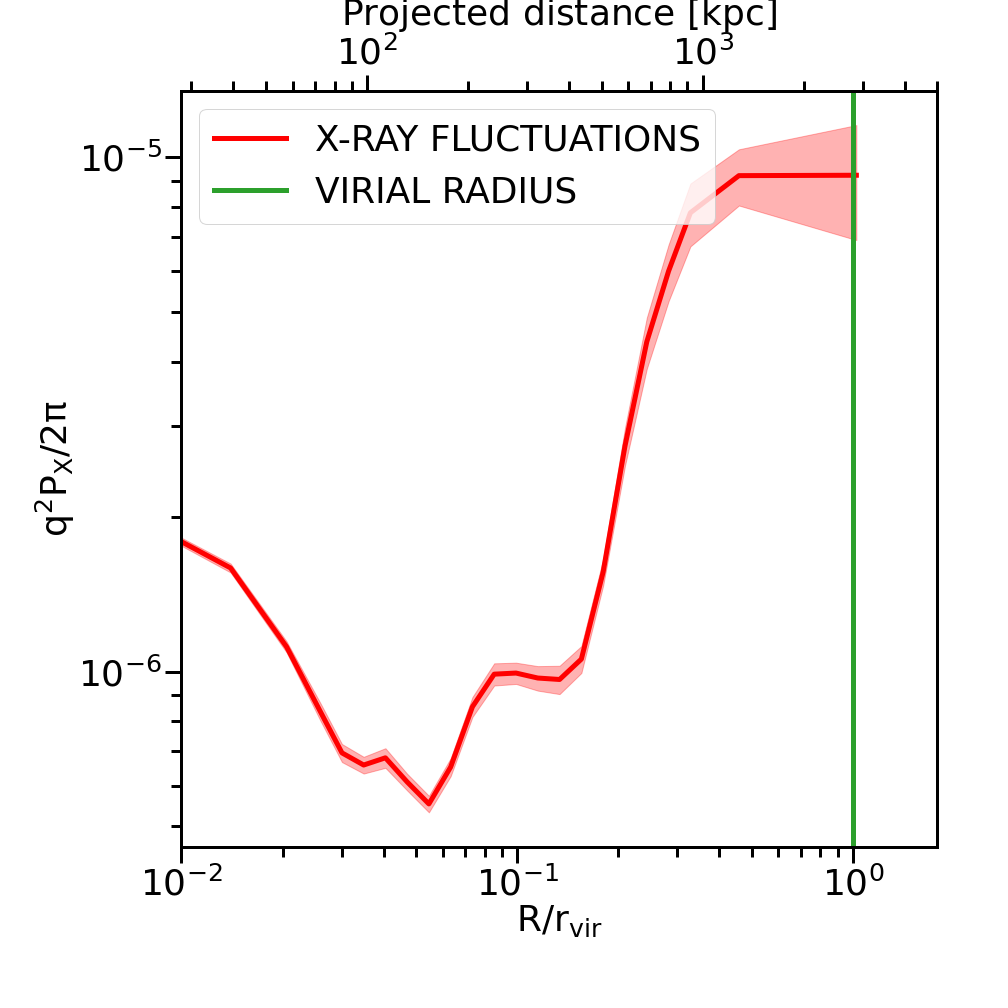}}
            \subfigure[]{\includegraphics[width=0.3\textwidth]{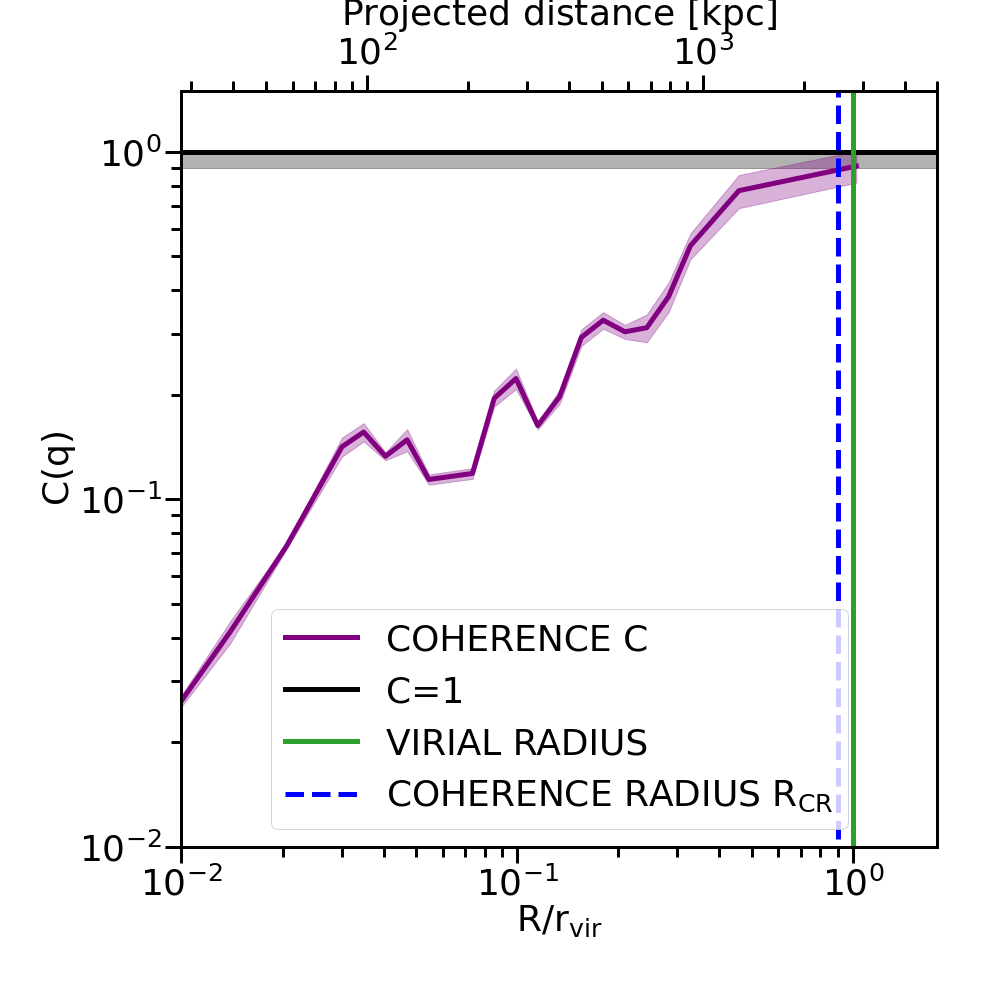}}
    \end{center}
    \caption{Analysis of the HSTFF cluster Abell 2744: (a) HST lensing map from the CATS collaboration; (b) Chandra X-ray surface brightness map; (c) computed mass dimensionless rms fluctuations; (d) computed X-ray dimensionless rms fluctuations; (e) computed gas-mass coherence as a function of scale. The bottom panels show the results as a function of the projected distance $R$ (top axis) and $R/r_{\rm vir}$ (bottom axis), with $r_{\rm vir}$ as virial radius.}
    \label{abell_2744}
\end{figure*}

\section{Observational and Simulated Datasets} \label{data}

\subsection{Mapping the mass distribution with gravitational lensing} \label{lensing}

The detailed spatial distribution of dark matter in galaxy clusters can be mapped though their observed gravitational lensing of distant background galaxies. Due to their large mass, galaxy clusters, as well as galaxies, locally deform space-time. Therefore, wave fronts of light emitted by a distant source travelling past a foreground galaxy cluster will be deflected and distorted. The deflection angle does not depend on the wavelength of light, as the effect is purely geometric. When distant background galaxies are in near perfect alignment with a massive foreground cluster, strong gravitational lensing effects are produced and the images of background galaxies are highly magnified, multiply imaged and their shapes are distorted, and typically gravitational giant arcs form. In the weak regime, when the alignment between observer, cluster and distant background galaxies is less perfect, the distortion induced by the cluster can only be detected statistically. Statistical methods are typically used to detect the resultant one percent or so change in the shape (ellipticity) of background galaxies. The observed quantities in cluster lensing studies are the magnitudes and shapes of the background population in the field of the cluster.
The cluster mass mapping involves the production of magnification maps, obtained through the solution of the lens equation for light rays originating from distant sources and deflected by the massive foreground cluster. The gravitational potential of the cluster is modeled as the 
sum of large- and small-scale mass contributions. The larger-scale contributions take into account the smooth larger-scale distribution of dark matter, often associated with the brightest central galaxies, and the small-scale components represent the masses of individual cluster member galaxies \citep{Natarajan_1997}. The lensing magnification ($\kappa$) and distortion ($\gamma$) are used to reconstruct the overall projected cluster mass (summing contributions from multiple scales) along the line of sight, permitting constraining the overall 2D gravitational potential. Photometric and spectroscopic data are also used to confirm cluster membership, as well as confirm the positions of lensed background sources. Determining cluster mass distributions from lensing observations is a rich and mature field, and the power of these techniques to reconstruct the mass distribution, by inverting the deflection map, has been amply demonstrated (see review by \cite{KneibNatarajan_2011}). 

Abell 2744 (redshift $z=0.308$) is one of the six galaxy clusters analyzed by the HSTFF program (\citealt{Lotz_2017}). The program focused on six deep fields centered on massive, highly efficient cluster lenses, which are preferentially actively merging systems. Several independent teams adopting a range of computational approaches were selected to produce a set of publicly available magnification and resultant lensing mass maps. For this proof of concept demonstration paper we started our investigation using the mass map from the team Clusters As TelescopeS (CATS; see \citealt{Natarajan_1997}, \citealt{Jullo_2009}, \citealt{Jauzac_2018} \citealt{Niemiec_2020} for further details). This model was our first choice because the results obtained with this technique has been demonstrated to be in good agreement with theoretical predictions from high resolution cosmological N-body simulations (\citealt{Natarajan_2017}) and from lens modelling comparison program conducted with a simulated cluster reported in \citealt{MeneghettiPNCoe_2017}. We then extended our coherence analysis to other independently derived mass models for the same cluster: the model from Bradac M. (\citealt{2005A&A...437...39B}, \citealt{2009ApJ...706.1201B}), from Diego J.M. (\citealt{10.1111/j.1365-2966.2005.09021.x}, \citealt{2007MNRAS.375..958D}), from GLAFIC team (\citealt{2010PASJ...62.1017O}, \citealt{Ishigaki_2015}, \citealt{Kawamata_2018}), from Williams L. (\citealt{2006MNRAS.367.1209L}, \citealt{Sebesta_2016}, \citealt{2017MNRAS.465.1030P}), and 3 different models from Zitrin A. (\citealt{Broadhurst_2005}, \citealt{Zitrin_2009}, \citealt{2013ApJ...762L..30Z}), namely the Light-Traces-Mass (LTM) method with a Gaussian smoothing, the LTM method with the mass map smoothed by a 2D Spline interpolation and, finally, the method adopting the Navarro-Frenk-White (NFW) dark matter distribution. From now on we will refer to these models as CATS, BRADAC, DIEGO, GLAFIC, WILLIAMS, ZITRIN LTM GAUSS, ZITRIN LTM, and ZITRIN NFW, respectively \footnote{The mass maps are publicly available on STScI's MAST archive: \dataset[10.17909/T9KK5N]{https://doi.org/10.17909/T9KK5N}}. The second cluster we study here is Abell 383 (redshift $z=0.189$), one of the 25 massive relaxed, cool core clusters studied in the HST CLASH program (\citealt{Postman_2012}). For this work, we first used the mass map for Abell 383 reconstructed by Zitrin with the NFW dark matter distribution (\citealt{Zitrin_2011}, \citealt{Zitrin_2012}), and then added the coherence analysis obtained with the other available models obtained using the LTM method\footnote{The mass maps are publicly available on STScI's MAST archive: \dataset[10.17909/T90W2B]{https://doi.org/10.17909/T90W2B}}. For both clusters, the entire available HST imaging data set was used, including observations with both the Advanced Camera for Surveys (ACS) and the Wide Field Camera 3 (WFC-3).

The CATS collaboration used LENSTOOL (\citealt{Jullo_2007}, \citealt{Jullo_2009}), an algorithm that combines observed strong- and weak-lensing data to constrain the cluster mass model. As noted above, the total mass distribution in a cluster was modelled as the linear sum of smooth, large-scale potentials, and perturbing mass distributions from individual cluster galaxies that are all modelled using Pseudo-Isothermal Elliptical Mass Distributions (PIEMDs) - a physically well motivated parametric form. A small-scale dark-matter clump is assigned to each major cluster galaxy, while a large-scale dark-matter clump represents the prominent concentration of cluster galaxies (see \citealt{Natarajan_1997,Jauzac_2016} for further details). The results of mass distributions obtained by this technique show good agreement with theoretical predictions from high-resolution cosmological N-body simulations (\citealt{Natarajan_2007}). Detailed comparison of three HSTFF clusters, including Abell 2744, with the Illustris cosmological suite of simulations can be found in \cite{Natarajan_2017}.

Abell 2744 shows an extraordinary amount of substructures, that are described in detail by \citealt{Jauzac_2016} and previously studied with prior HST shallower data, for instance, by \citealt{Merten_2011} and \citealt{Medezinski_2016}. All the analysis report that Abell 2744 is undoubtedly undergoing a complicated merging process, highlighted by the separation between the distribution of mass (panel (a) of Fig.\ref{abell_2744}) and baryonic components (panel (b) of Fig. \ref{abell_2744}). The largest structure in the inner parts, referred to as the Core, is visible in the center of panel (a) of Fig. \ref{abell_2744}. The center of the cluster was chosen to match the location of the Core's Brightest Cluster Galaxy (BCG), i.e. $(3\degree.59,-30\degree.40)$ (J2000.0). 

\begin{figure*}
    \begin{center}
            \subfigure[]{\includegraphics[width=0.352\textwidth]{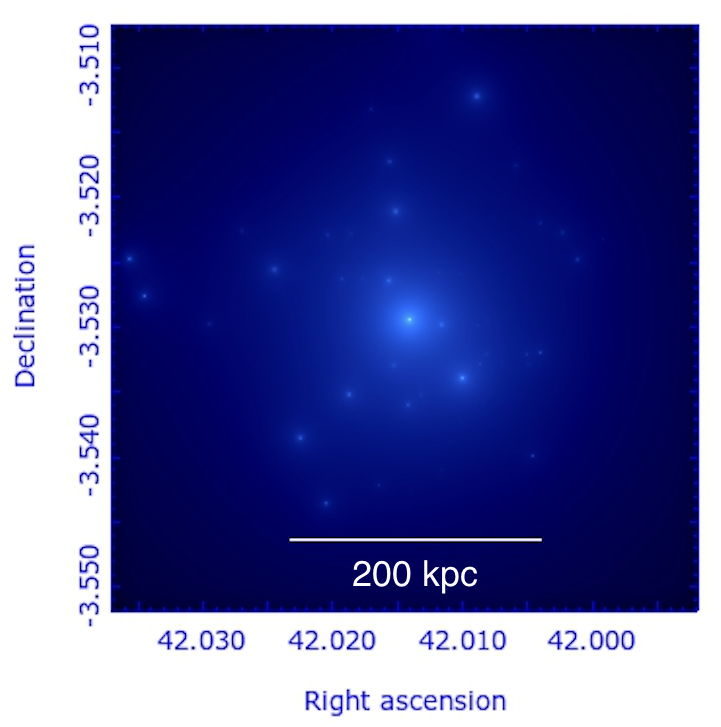}}
            \subfigure[]{\includegraphics[width=0.35\textwidth]{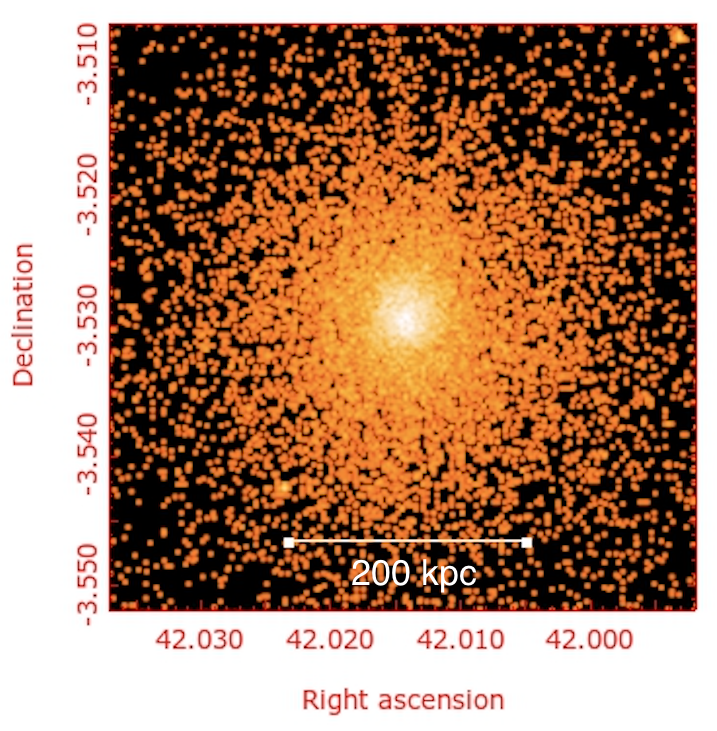}}  
            \subfigure[]{\includegraphics[width=0.3\textwidth]{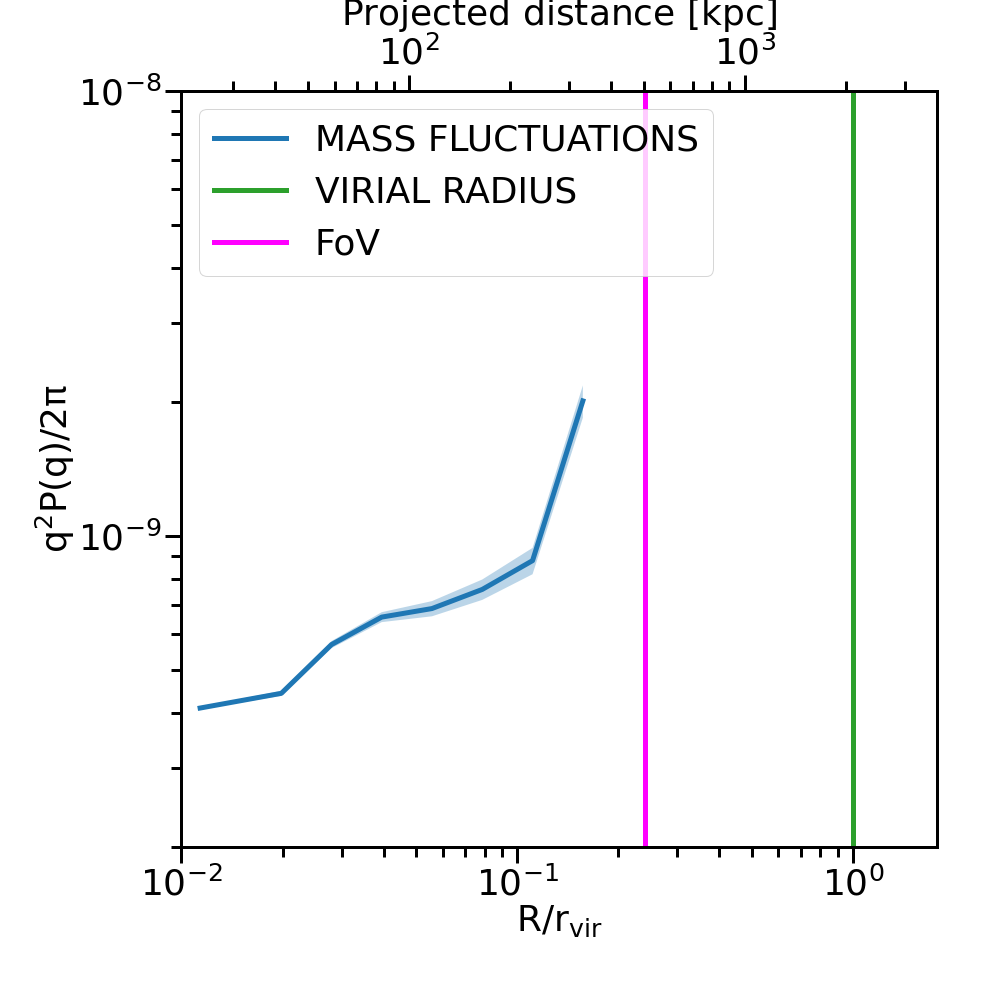}} 
            \subfigure[]{\includegraphics[width=0.3\textwidth]{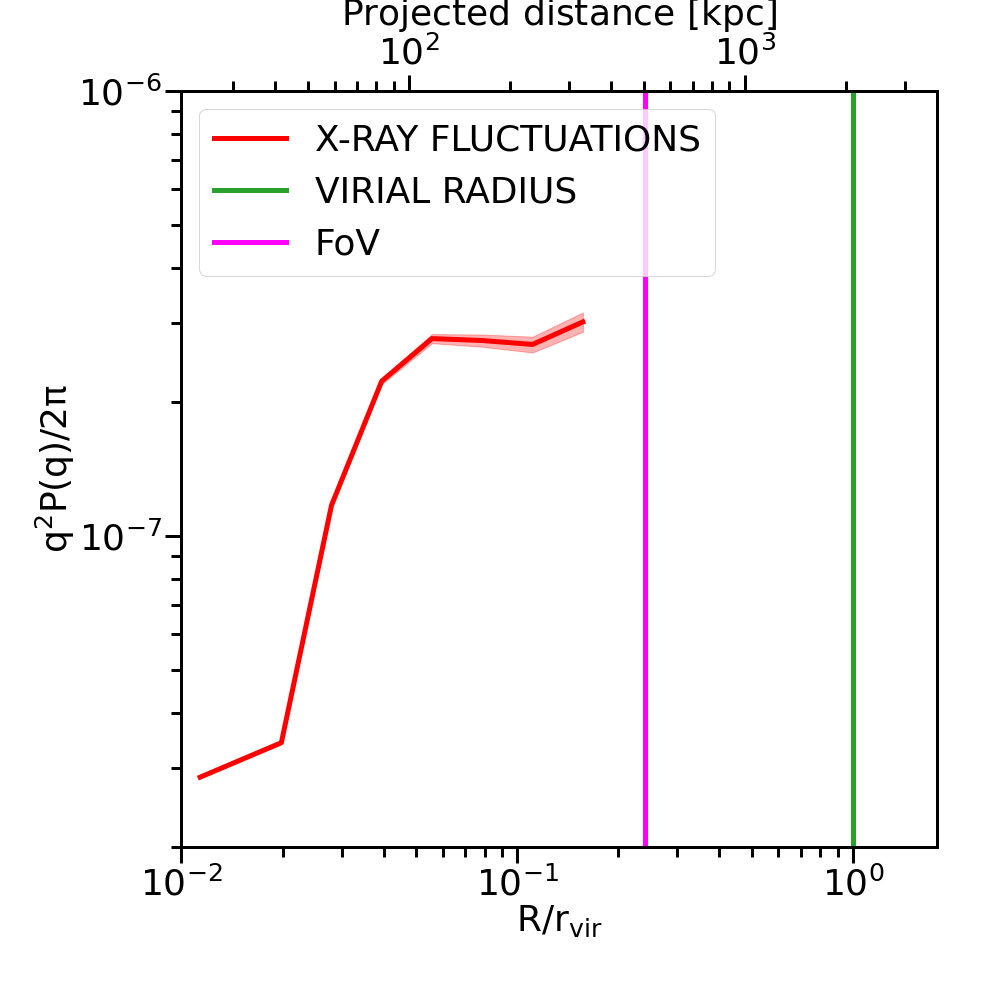}}
            \subfigure[]{\includegraphics[width=0.3\textwidth]{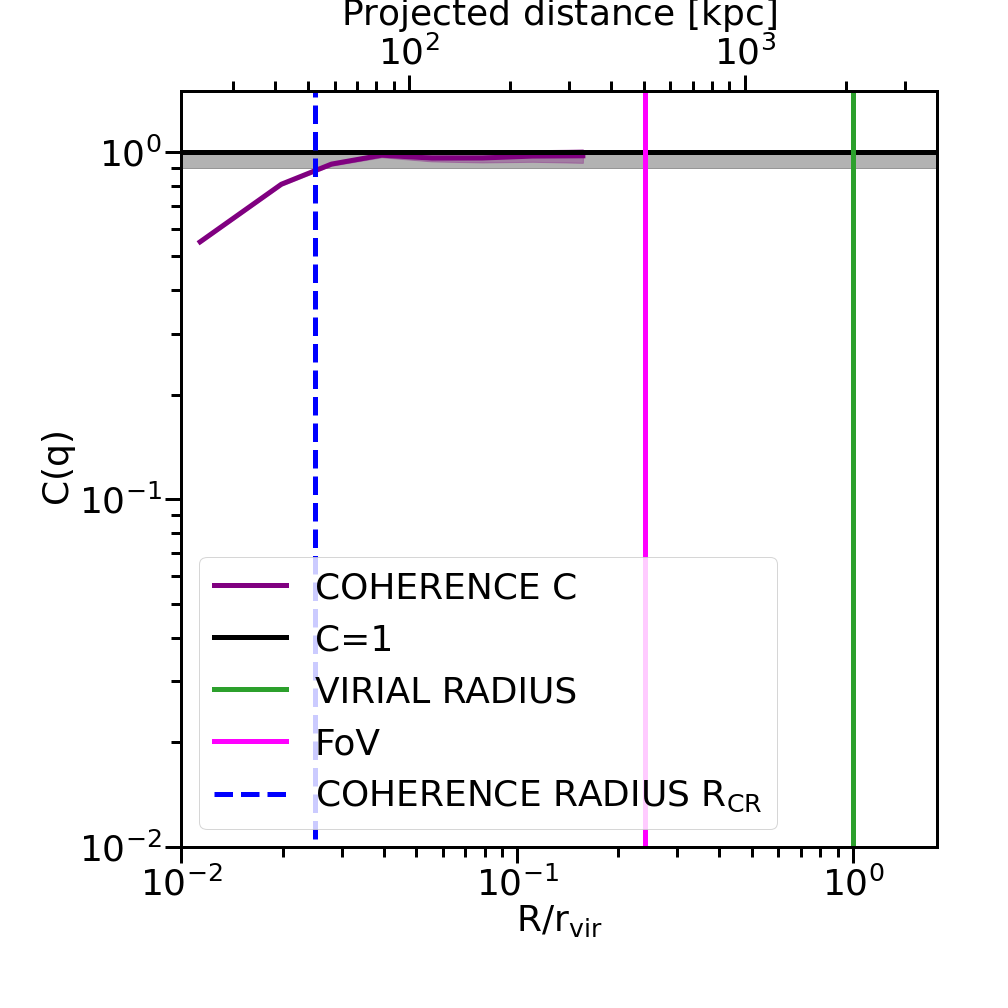}}
    \end{center}
    \caption{Analysis of the CLASH cluster Abell 383: (a) CLASH lensing map of Abell 383; (b) Chandra X-ray surface brightness map; (c) computed mass dimensionless rms fluctuations; (d) computed X-ray dimensionless rms fluctuations; (e) computed gas-mass coherence as a function of scale. The bottom panels show the results as a function of the projected distance $R$ (top axis) and $R/r_{\rm vir}$ (bottom axis), with $r_{\rm vir}$ as virial radius.}
    \label{abell_383}
\end{figure*}

Abell 383, on the other hand, does not show the complexity and clumpiness exhibited by Abell 2744, and the dark matter and the baryonic gas appear to trace each other well (the mass map and X-ray distribution are shown in panel (a) and (b) of Fig. \ref{abell_383}, respectively). Indeed, \citealt{Zitrin_2011} showed that the overall mass profile is well fitted by a Navarro-Frenk-White (NFW) profile with a virial mass $M_{\rm vir}=5.37^{+0.70}_{-0.63} \pm 0.26 \times 10^{14}M_{\odot}$ and a relatively high concentration parameter $c=8.77^{+0.44}_{-0.42} \pm 0.23 $. Combining weak and strong lensing, the mass distribution in the inner regions of the relaxed galaxy cluster Abell 383 can be reconstructed using a parametric model that consists of PIEMDs for the individual cluster galaxies, and elliptical Navarro-Frenk-White (eNFW) halo to map the smoother larger-scale dark-matter distribution, whose center was fixed at the center of the BCG at coordinates $(42\degree.01,-3\degree.53)$ (J2000.0).

For both clusters authors adopted the $\Lambda$CDM concordance cosmology model with $\Omega_m=0.3$, $\Omega_{\Lambda}=0.7$ and a Hubble constant $H_0=70\,km\,s^{-1}{\rm Mpc}^{-1}$. With these values of the cosmological parameters, for Abell 2744, at redshift $z=0.308$, the corresponding luminosity distance is $1600\,{\rm Mpc}$, while for Abell 383, at redshift $z=0.187$, the corresponding luminosity distance is $909\,{\rm Mpc}$; for the former cluster 1" corresponds to $4.54\,{\rm kpc}$, while for the latter 1" translates to $3.15\,{\rm kpc}$. The mass maps have a Field of View (FoV) of about $4.5 \times 4.5\,{\rm Mpc}$ and $0.5 \times 0.5\,{\rm Mpc}$, respectively.

Multiple teams deploying independent mass modeling codes were selected to produce models for the HSTFF cluster sample including Abell 2744. The BRADAC, DIEGO and WILLIAMS teams used non-parametric techniques to derive their mass models, while the CATS, GLAFIC \& ZITRIN teams adopted a parametric approach. The parametric models start with a prior that includes parametric analytic functional forms for the density profiles of multiple mass components including larger scale and smaller galaxy scale halos that are assumed to comprise the total mass distribution \cite{Natarajan_1997}. In addition, empirical relations that couple the mass to the light of cluster member galaxies hosted by the smaller scale subhalos are assumed. The mass model parameters are optimized to reproduce observed multiple image properties. Both the non-parametric BRADAC model, as well as the parametric CATS team model combine strong- and weak-lensing information (see \citealt{2005A&A...437...39B} and \citealt{2009ApJ...706.1201B}) that permits breaking the mass-sheet degeneracy (\citealt{1995A&A...294..411S}). The DIEGO team model is based on their own independently constructed Weak and Strong Lensing Analysis Package (WSLAP; see \citealt{10.1111/j.1365-2966.2005.09021.x} and \citealt{2007MNRAS.375..958D} for more information), that permits derivation of lens models based on strong and, when available, weak lensing data. However, only strong lensing data were used for the mass reconstruction of Abell 2744. With this package the mass distribution is built as a superposition of Gaussian functions and a compact component that traces the light of the member galaxies, modeled as the observed light around prominent member galaxies. The non-parametric WILLIAMS model uses their GRALE code (\citealt{2006MNRAS.367.1209L}, \citealt{Sebesta_2016}, \citealt{2017MNRAS.465.1030P}), a flexible, form-free strong lens reconstruction method that uses an adaptive grid and no prior information about the cluster member galaxies. From an initial coarse grid, populated with a basis set, such as projected Plummer density profiles, a genetic algorithm is used to iteratively refine the mass map solution and, at every iteration, the dense regions are resolved with a progressively finer grid. The final map is given by a superposition of a mass sheet and many Plummer profiles, each with its own size and weight, that are determined by the genetic algorithm. As for the remaining parametric models, GLAFIC is based on the homonyim software package for analyzing gravitational lensing (see \citealt{2010PASJ...62.1017O}, \citealt{Ishigaki_2015} and \citealt{Kawamata_2018} for more details), and the ZITRIN models adopt the Light Traces Mass (LTM) method (\citealt{Broadhurst_2005}, \citealt{Zitrin_2009}). In the latter, the LTM assumption is adopted for both the galaxies and the dark matter components. The first component of the model is given by the superposition of all the galaxy contributions, determined by the identification of the cluster members. Cluster members are chosen from the identified red-sequence and are assigned a power-law mass density profile that is in turn scaled by the galaxy's luminosity. The mass map is then smoothed with a 2D Spline interpolation (LTM method) or a Gaussian kernel (LTM GAUSS method), to obtain a smooth component representing the DM mass density distribution. Finally, the two mass components are added with a relative weight, and a 2-component external shear component is included to allow for additional flexibility in order to better reproduce the ellipticity of the critical curves.

\subsection{X-ray data} \label{xray}

For the X-ray images, we used the Chandra ACIS-I data in the 0.5-7.0 keV energy band. The dataset for Abell 2744 was obtained from 3 pointings, with a total exposure time of 92.34 ks (OBSIDs: 7915, 8477 and 8557), while for Abell 383 (OBSIDs: 2320 and 524) we used 2 pointings, with a total exposure of 29.25 ks. While XMM-Newton data could have also been used for this work, with its on-axis HPD angular resolution of $\sim 15"$, it is not ideal as the broad PSF blurs the smallest-scale structures and the required excision of foreground sources seriously limits the amount of area available for the analysis presented here. The Chandra data, on the other hand, with an on-axis HPD angular resolution of $\sim 0.5"$, allows us to resolve down to few kpc scales, at the redshifts of these two clusters. Indeed, for Abell 2744, located at $z=0.308$, $0.5"$ and $15"$ correspond to $\sim 2.3\,{\rm kpc}$ and $\sim 68.1\,{\rm kpc}$, respectively. In future follow-up work, we plan to enlarge our analysis sample and study clusters over a wider range of redshifts, up to $z=0.89$. Consequently, Chandra archival data will be ideal for our purpose.

To compute the maps for the power spectrum and coherence analysis, events were sorted in arrival time to create odd and even images, that we refer to as $A$ and $B$ subsets. These two subsets were used to create the signal and noise maps $1/2(A+B)$ and $1/2(A-B)$, respectively (both divided by the exposure maps). Indeed, following this approach, we obtain two identical images, with the same exposure, with half the photons each. Therefore, the two images contain the same information in terms of the signal but they have their own noise, and any systematic error would manifest itself very similarly in the $A$ and $B$ images. Consequently, the $1/2(A-B)$ difference images provide a useful means of characterizing the Poisson fluctuation amplitude (the A-B method was previously used, for instance, by \citealt{Kashlinsky2005}, \citealt{Cappelluti_2013} and \citealt{Cappelluti_2017}, \citealt{Li_2018}). 

\begin{figure}
\includegraphics[width=1.0\linewidth]{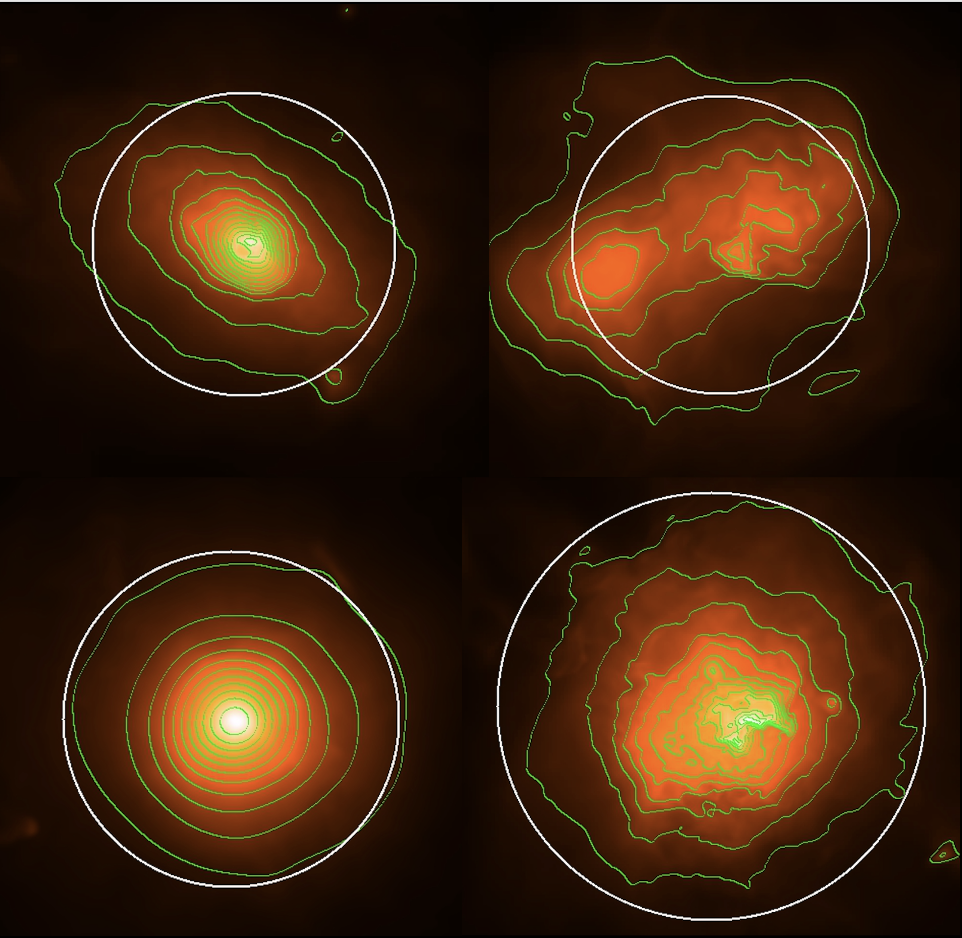}
\caption{Simulated cluster analogs: X-ray images of the four OMEGA500 clusters chosen as analogs for Abell 2744 and Abell 383. The green contours clearly reveal different dynamical states for these simulated clusters. The halo C-3 in the lower left corner is the only cluster that appear to be well relaxed. The most unrelaxed cluster is the halo C-2 in the upper right corner, whose contours highlight the presence of an on-going merger. The white circles correspond to $R_{500}$: $0.81\,{\rm Mpc}$ for the halo C-1 in the upper left corner, $0.82\,{\rm Mpc}$ for the halo C-2 in the upper right corner, $0.90\,{\rm Mpc}$ for the halo C-3 in the lower left corner and $1.22\,{\rm Mpc}$ for the halo C-4 in the lower right corner.}
\label{contours}
\end{figure}

Each X-ray map was masked for detected point sources, whose identification was conducted cross-checking the list we obtained from the Chandra Source Catalog with that obtained running the CIAO tool \textit{wavdetect}. A total of 52 and 2 X-ray point sources were erased in the FoVs of Abell 2744 and Abell 383, respectively (we want to highlight that the discrepancy in the number of identified X-ray sources is due to the fact that, while for Abell 2744 the mass map was reconstructed for the entire cluster, for Abell 383 we only have the map reconstruction of the cluster core, corresponding to a projected area of $0.5 \times 0.5\,{\rm Mpc} $, as highlighted also at the end of this subsection). The masked maps were obtained by cutting circular regions of $7"$ radius, similarly to what was done in previous Chandra analysis of diffuse X-ray components. For instance, \citealt{Cappelluti_2012} and \citealt{Cappelluti_2017} chose to remove circular regions with radii of $5"$ and $7"$, respectively. We conducted our analysis using masks with both these values and did not obtain significant changes in the power spectrum and coherence results (in both cases the masked regions do not exceed $\sim 2\%$ of the total area). In addition, in order to take into account the effects of sensitivity variation across the field of view, every pixel \textit{i} of the count maps was weighted by a factor \textit{$E(i)/<E>$}, where \textit{$E(i)$} is the exposure at the pixel \textit{i} and \textit{$<E>$} is the mean exposure in the field (this method was previously tested and adopted by \citealt{Cappelluti_2013}). The masked X-ray maps were then divided by the exposure maps and cleaned to remove instrumental and cosmic foregrounds through the subtraction of the background maps from the two $A$ and $B$ subsets. 

All the archival data were calibrated using the software package CIAO. We processed level 1 data products  with the CIAO tool \textit{chandra\textunderscore repro}, retaining only valid event grades, and we examined the light curves of each individual observation for flares. The particle background was subtracted by tailoring images taken by ACIS-I in stowed mode, in the same energy band under analysis. Indeed, when ACIS is not exposed to the sky, the stowed image only contains events due to particles. Then, the stowed image must be renormalized to match the actual background level in the observations, since the background level is not constant in time. Following the recipe proposed by \citealt{Hickox_2006}, the stowed images in the band 0.5-7.0 keV were scaled by the ratio between the total counts measured in the real image and those measured in the stowed image, both in the band 9.5-12 keV, an energy range in which Chandra does not have counts of astrophysical origin but only particle events (this approach is justified by the fact that the shape of the particle background spectrum is constant in time but its amplitude is not). 

The masked, background-subtracted, exposure-corrected images were merged and mosaic images were created using the python \textit{reproject.mosaicking} sub-package. Since the computation of the cross-power spectrum and, consequently, the coherence, requires the same angular binning, we used the same sub-package \textit{reproject.mosaicking} to re-grid the images and match the pixel resolution of the mass maps (the photon counts of the final images were multiplied by the ratio of the final and initial pixel areas): $1000\times 1000\  1"$ pixels, for a total FoV of $1000\times 1000\,{\rm arcsec}$ for Abell 2744, corresponding to $4.5\times 4.5\,{\rm Mpc}$, and $2500\times 2500\ 0.065"$ pixels, for a total FoV of $162\times 162\,{\rm arcsec}$ for Abell 383, corresponding to $0.5 \times 0.5\,{\rm Mpc} $.

\begin{figure*}
    \begin{center}
            \subfigure[]{\includegraphics[width=0.35\textwidth]{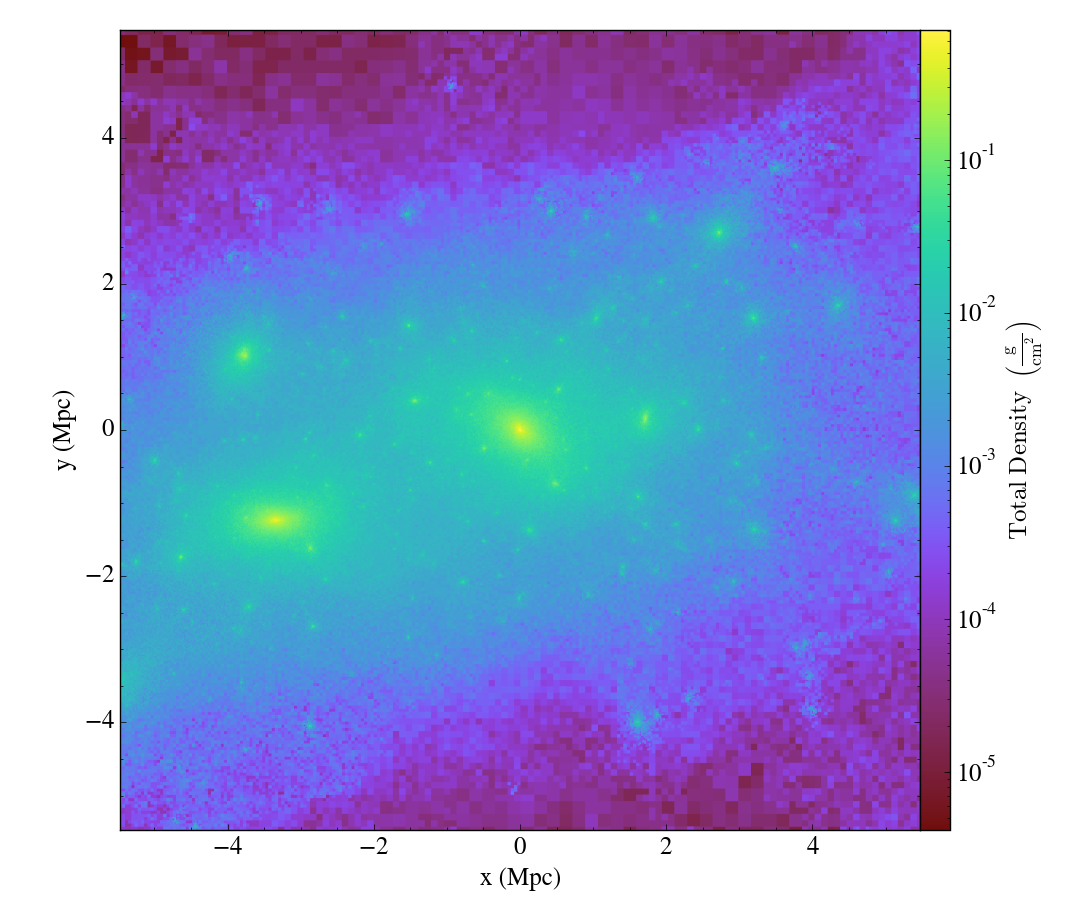}}
            \subfigure[]{\includegraphics[width=0.35\textwidth]{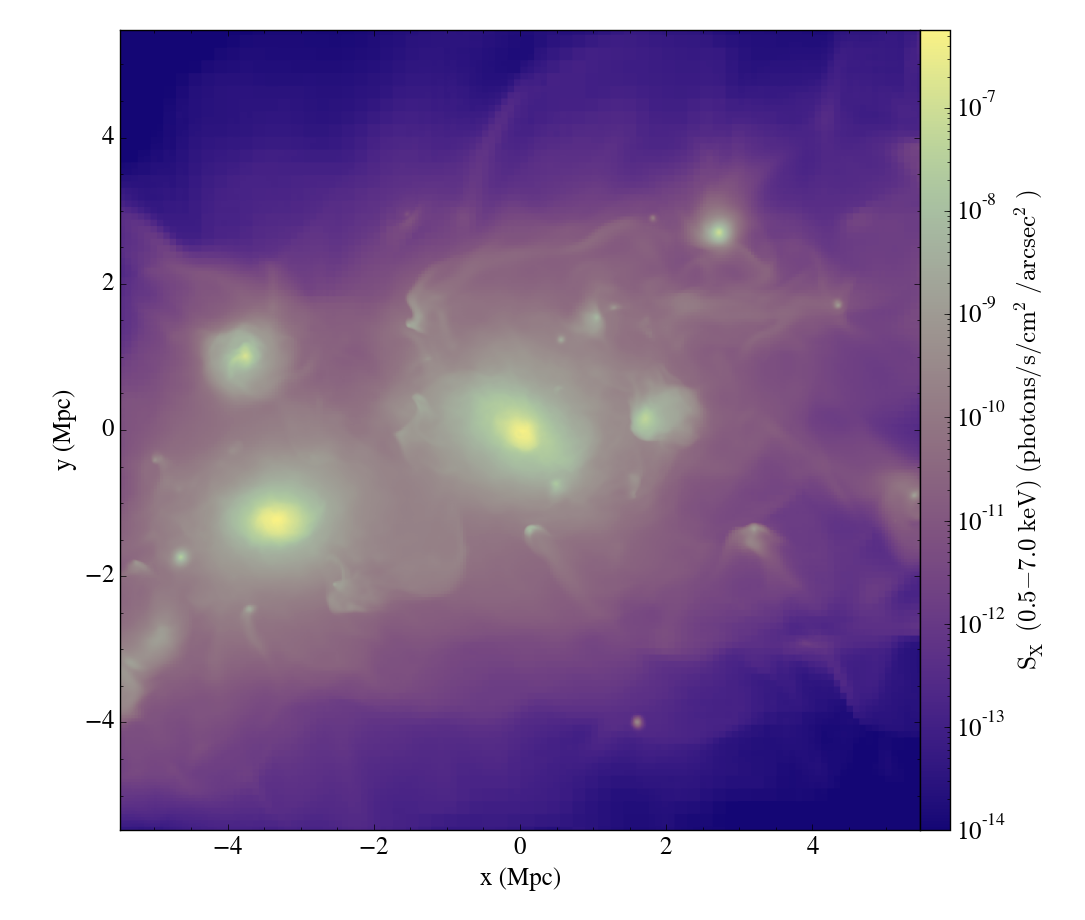}}        
            \subfigure[]{\includegraphics[width=0.3\textwidth]{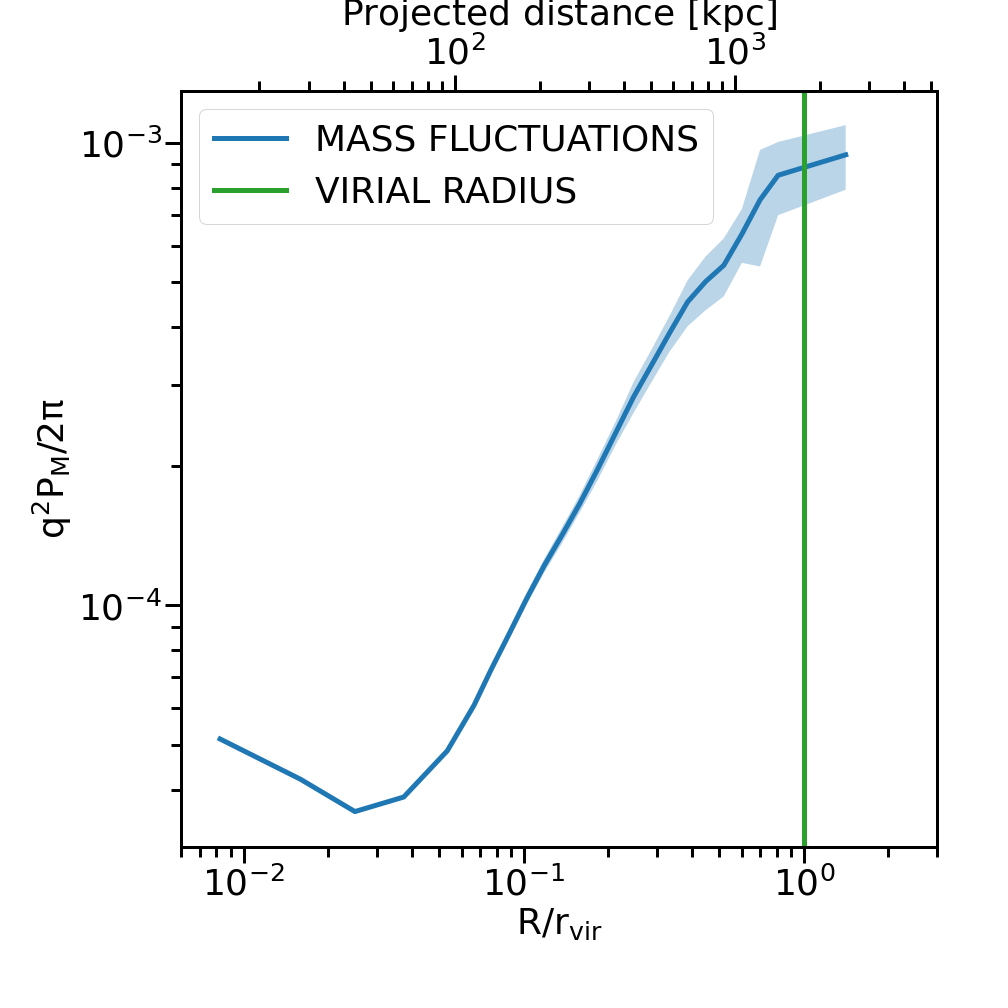}} 
            \subfigure[]{\includegraphics[width=0.3\textwidth]{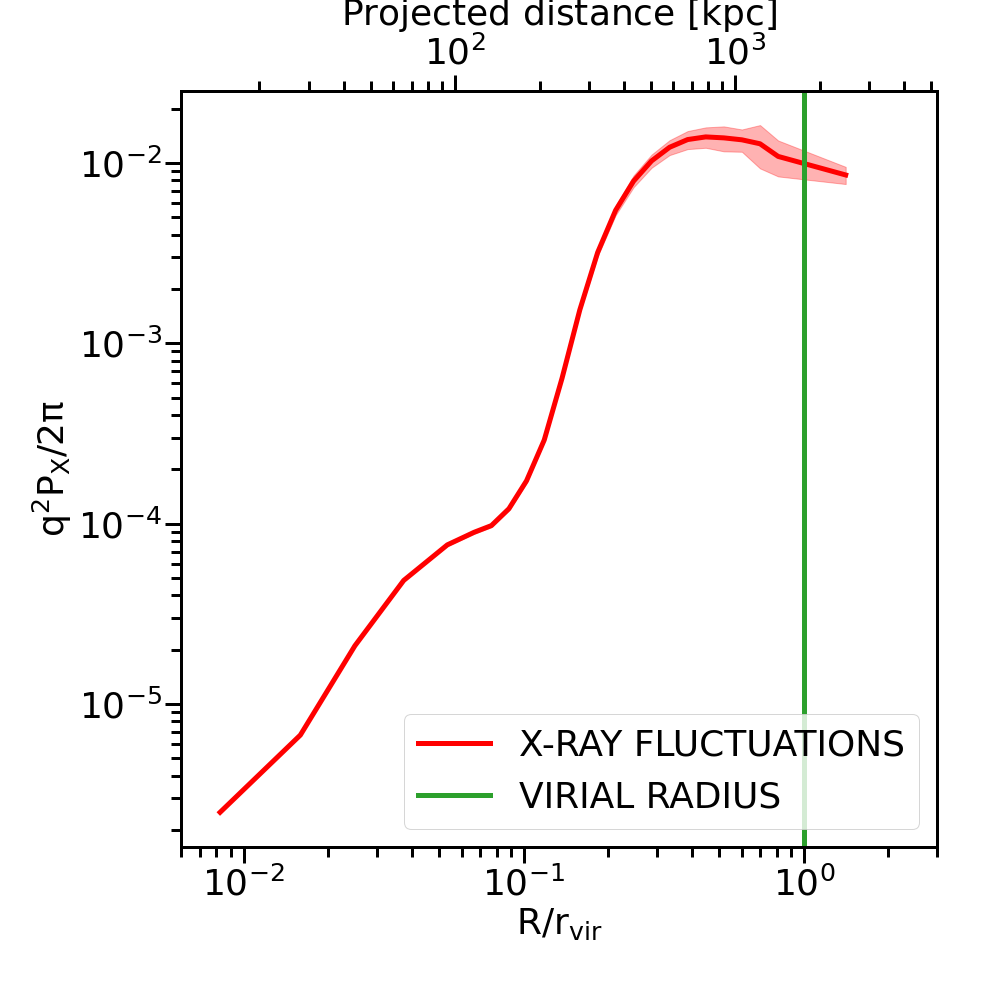}}
            \subfigure[]{\includegraphics[width=0.3\textwidth]{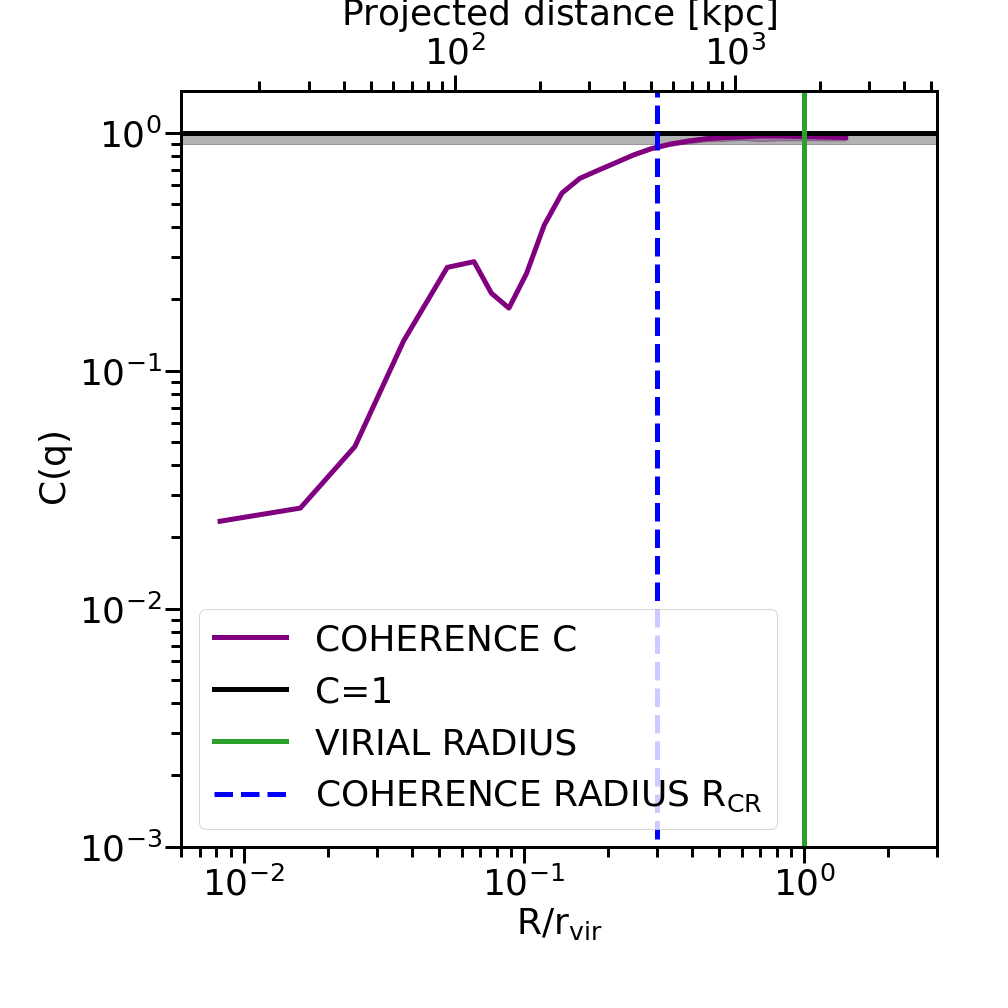}}
    \end{center}
    \caption{Analysis of the Simulated Cluster Analog C-1: (a) mass map of the simulated cluster halo C-1; (b) X-ray surface brightness map of halo C-1; (c) computed mass dimensionless rms fluctuations of halo C-1; (d) computed X-ray dimensionless rms fluctuations of halo C-1; (e) computed gas-mass coherence of halo C-1 as a function of scale. The bottom panels show the results as a function of the projected distance $R$ (top axis) and $R/r_{\rm vir}$ (bottom axis), with $r_{\rm vir}$ as virial radius.}
    \label{halo1}
\end{figure*}

\subsection{Simulated clusters from the OMEGA500 simulations} \label{omega500}

The OMEGA500 simulations consist of a sample of $z=0$ galaxy clusters obtained from large high-resolution hydrodynamical cosmological simulations, performed using the Adaptive Refinement Tree (ART) N-body+gas-dynamics code, in a flat $\Lambda$CDM model with WMAP five-year (\textit{WMAP5}) cosmological parameters $\Omega_m=0.27$, $\Omega_{\Lambda}=0.73$, $H_0=70\,{\rm km\,s^{-1} Mpc^{-1}}$ (see \citealt{1989JCoPh..82...64B}, \citealt{Kravtsov_1997}, \citealt{Rudd_2008}, \citealt{Nelson_2014} for further details). The sample offers a wide range of masses and dynamical states. We note that the difference between the cosmological parameters adopted for these simulations and those chosen for the lensing maps does not affect the structures on the scales we analyze in this work. The halo abundance is the most sensitive relevant quantity that depends strongly on cosmological parameters (\citealt{1974ApJ...187..425P}) but, as our analysis is focused on individual halos, this small difference in the adopted values for $\Omega_m$ and  $\Omega_{\Lambda}$ is not impactful.

The code adopted for these cosmological simulations uses adaptive refinement in space and time and non-adaptive refinement in mass, in order to achieve the dynamic ranges to resolve the cores of halos. Starting from the standard assumed high-redshift initial conditions for matter fluctuations, derived from a physically motivated power spectrum, the dynamic evolution of gas, dark matter, and stars are all followed in an expanding cosmological background. Clusters of galaxies form in these simulations at the intersection between filaments of gas and dark matter. The co-moving length of the simulated box for this simulation suite is $500\,h^{-1}$\,Mpc, with a maximum co-moving spatial resolution of $3.8\,h^{-1}\,{\rm kpc}$. In the co-moving box used in our analysis, projected quantities from the regions surrounding the clusters were taken. In particular, projections were taken along the three major axes of the simulation domain: along the $x$, $y$, and $z$ directions. In this investigation, we used the z-projection of the total mass density and X-ray emissivity. In this paper, we present the proof of concept of our new proposed technique using two observed clusters and their mass matched simulated analogs. In future follow-up studies, we plan to delve deeper into the assessing detailed properties of the simulated clusters including taking projections along different lines of sight that might be useful to infer the 3D power spectrum and to also investigate the role of projection effects on the derived power spectra. We also plan to compare in detail simulated clusters from different cosmological simulation suites that adopt independent prescriptions for galaxy formation and feedback. These implemented physical processes may imprint signatures on the power spectrum - these will be studied in future work. 

All images used in this analysis are centered on the cluster potential minimum. For the comparison with Abell 2744 and Abell 383, we selected 4 simulated clusters, whose virial mass, virial radius and dynamical state are listed in Table \ref{table:1}. From cluster mass halos in the simulations, we chose 4 targets with different masses and dynamical states, determined via the morphology of their X-ray surface brightness emission contours. We made this choice to include a wider range of simulated clusters that span a range of evolutionary stages. Below, we describe how we determined the different dynamical states of the targets investigated in this work. 

One of the methods adopted to classify the dynamical state of galaxy clusters is based on their X-ray morphologies (\citealt{2015MNRAS.449..199M}, \citealt{2016MNRAS.455.2936S}). Similarly as reported in \citealt{2016MNRAS.455.2936S}, the determination of the dynamical state is usually based on the following criteria as a measure of relaxedness: 1) the presence in the X-ray image of a single, distinguished cluster core with a small displacement with respect to the bulk of the ICM; 2) the shape of contours, with round and elliptical contours suggesting that the cluster core is relaxed; 3) the presence of substructures or disturbances in the ICM between the cluster core region and $R_{500}$. Fig.~\ref{contours} shows a zoomed-in view of the X-ray maps of the simulated halos C-1 through C-4, with X-ray contours shown in green. The white circle corresponds to $R_{500}$. The halo C-1, in the upper left corner of Fig. \ref{contours}, meets the first and the second criteria, revealing a single cluster core with elliptical contours, but the presence of tiny disturbances within $R_{500}$ suggests that it is not fully relaxed. The halo C-2, in the upper right corner of Fig. \ref{contours}, is completely unrelaxed, since it shows the presence of two substructures, indicating an actively on-going merger. The halo C-3, in the lower left corner of Fig. \ref{contours} is the only galaxy cluster in our simulated sample that shows a good level of relaxation, with a distinguished core, smooth almost-round contours and no disturbances or substructures within $R_{500}$. Finally, the halo C-4, in the lower right corner of Fig. \ref{contours}, seems to be highly unrelaxed, since the irregular contours highlight the presence of a double peak at the center. 

\begin{figure*}
    \begin{center}
            \subfigure[]{\includegraphics[width=0.35\textwidth]{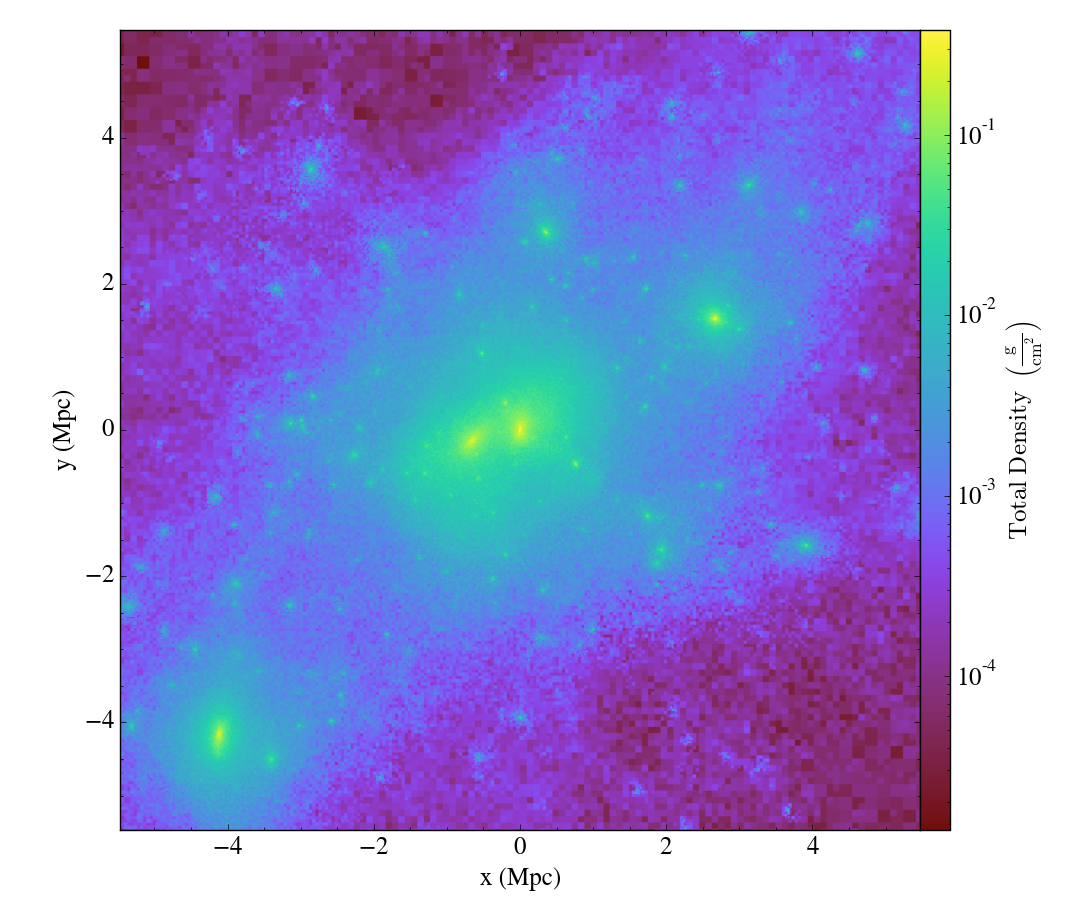}}
            \subfigure[]{\includegraphics[width=0.35\textwidth]{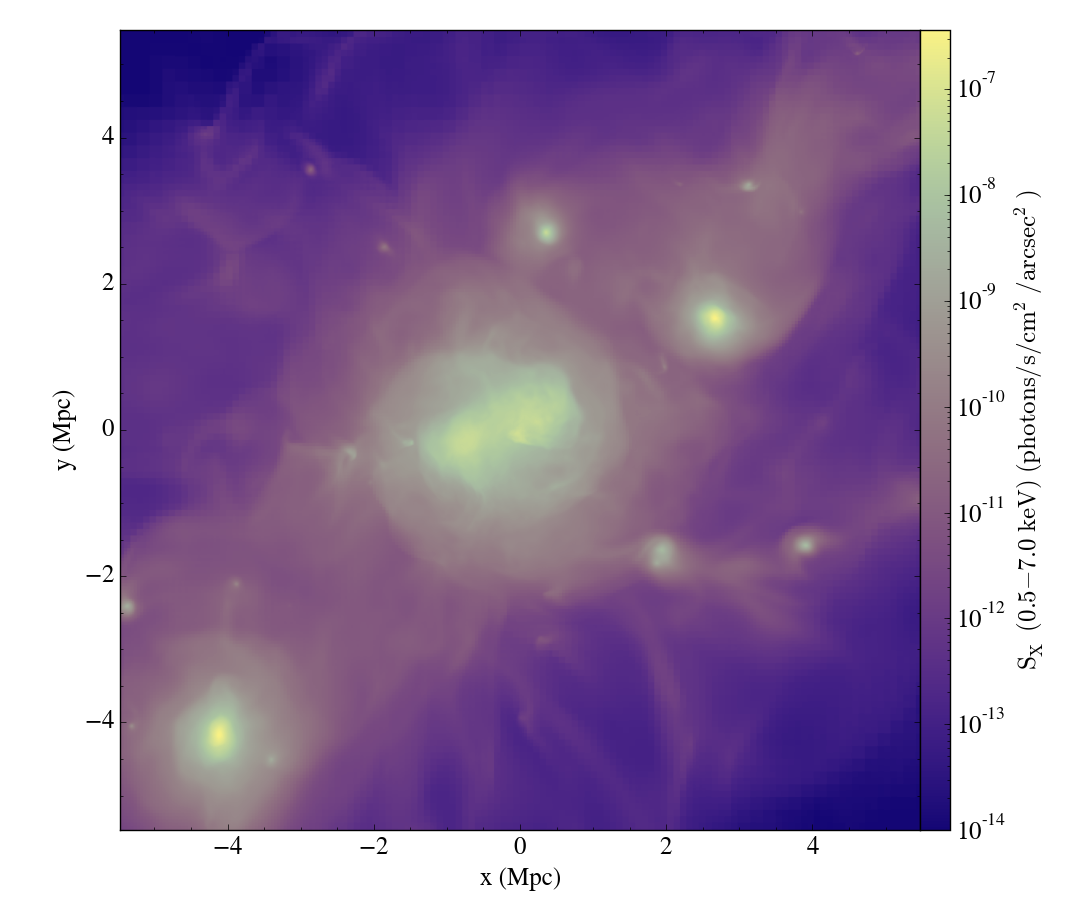}}        
            \subfigure[]{\includegraphics[width=0.3\textwidth]{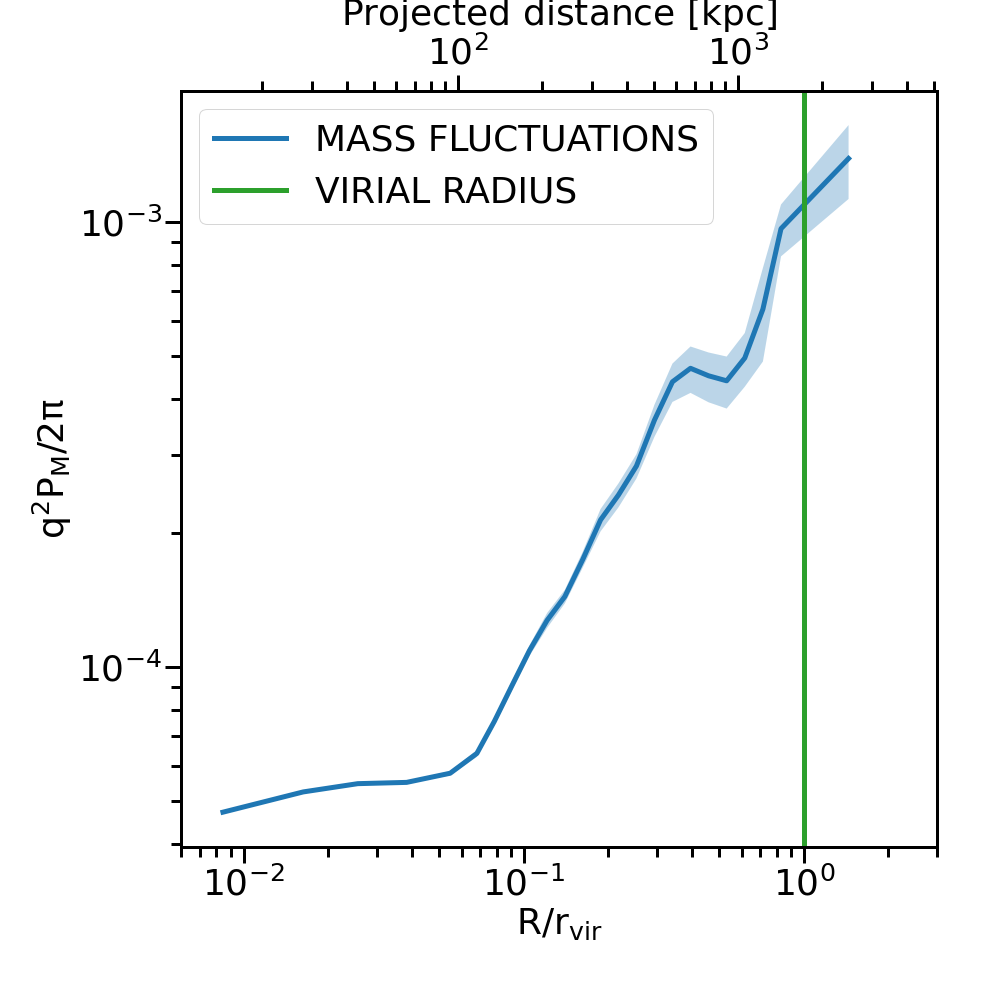}} 
            \subfigure[]{\includegraphics[width=0.3\textwidth]{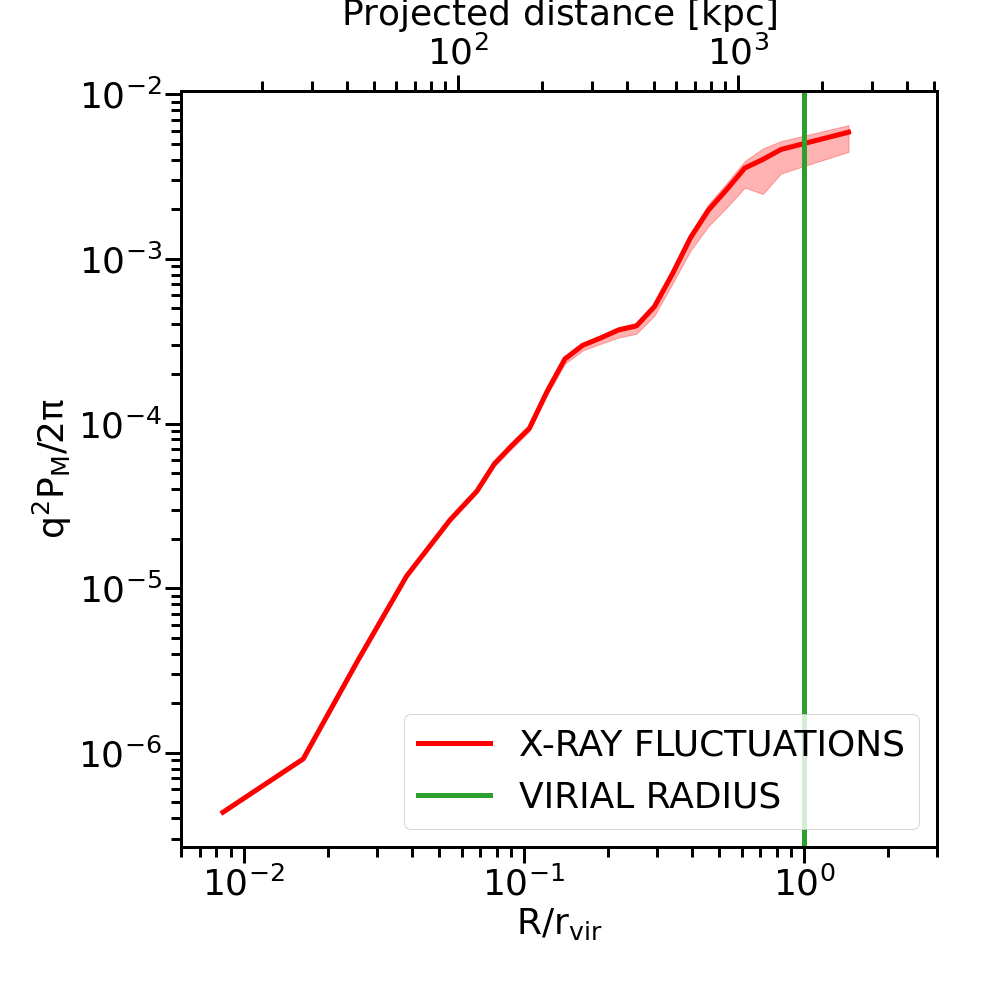}}
            \subfigure[]{\includegraphics[width=0.3\textwidth]{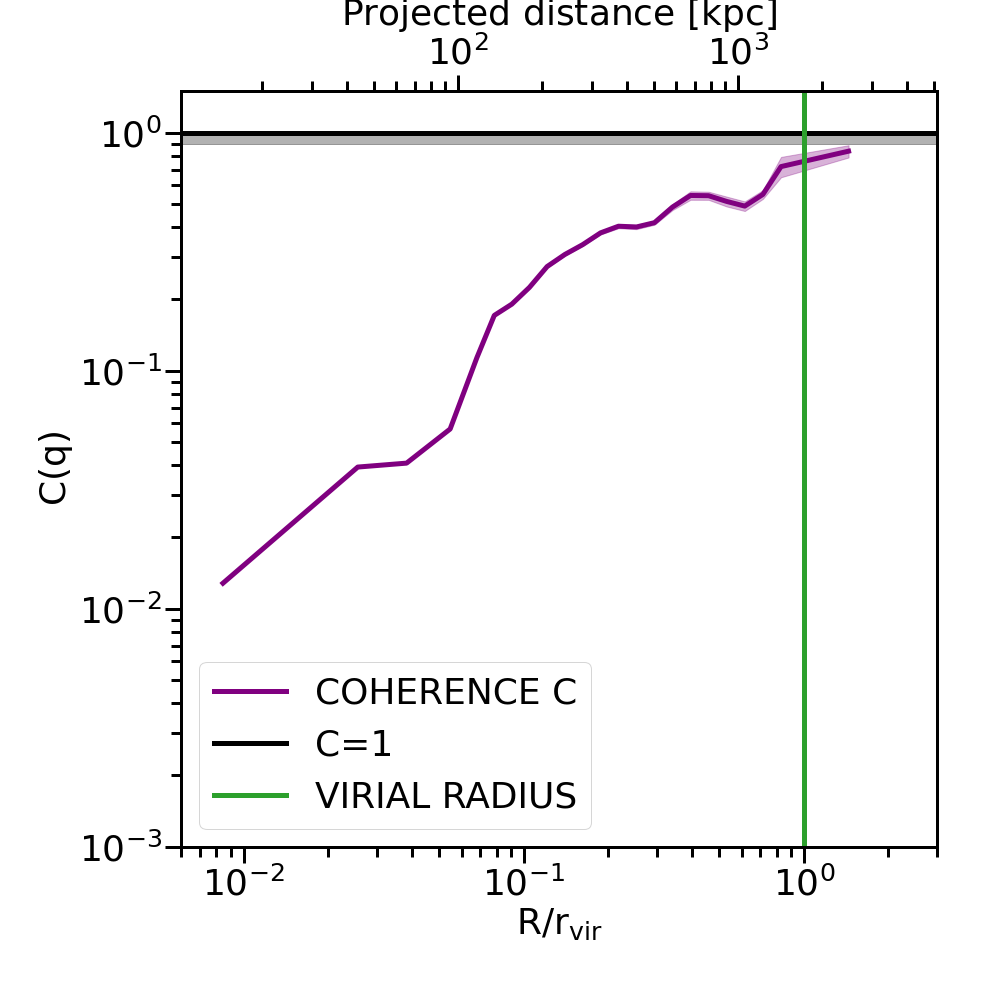}}
    \end{center}
    \caption{Analysis of the Simulated Cluster Analog C-2: (a) mass map of the simulated cluster halo C-2; (b) X-ray surface brightness map of halo C-2;(c) computed mass dimensionless rms fluctuations of halo C-2; (d) computed X-ray dimensionless rms fluctuations of halo C-2; (e) computed gas-mass coherence of halo C-2 as a function of scale. The bottom panels show the results as a function of the projected distance $R$ (top axis) and $R/r_{\rm vir}$ (bottom axis), with $r_{\rm vir}$ as virial radius.}
    \label{halo2}
\end{figure*}

\begin{figure*}
    \begin{center}
            \subfigure[]{\includegraphics[width=0.35\textwidth]{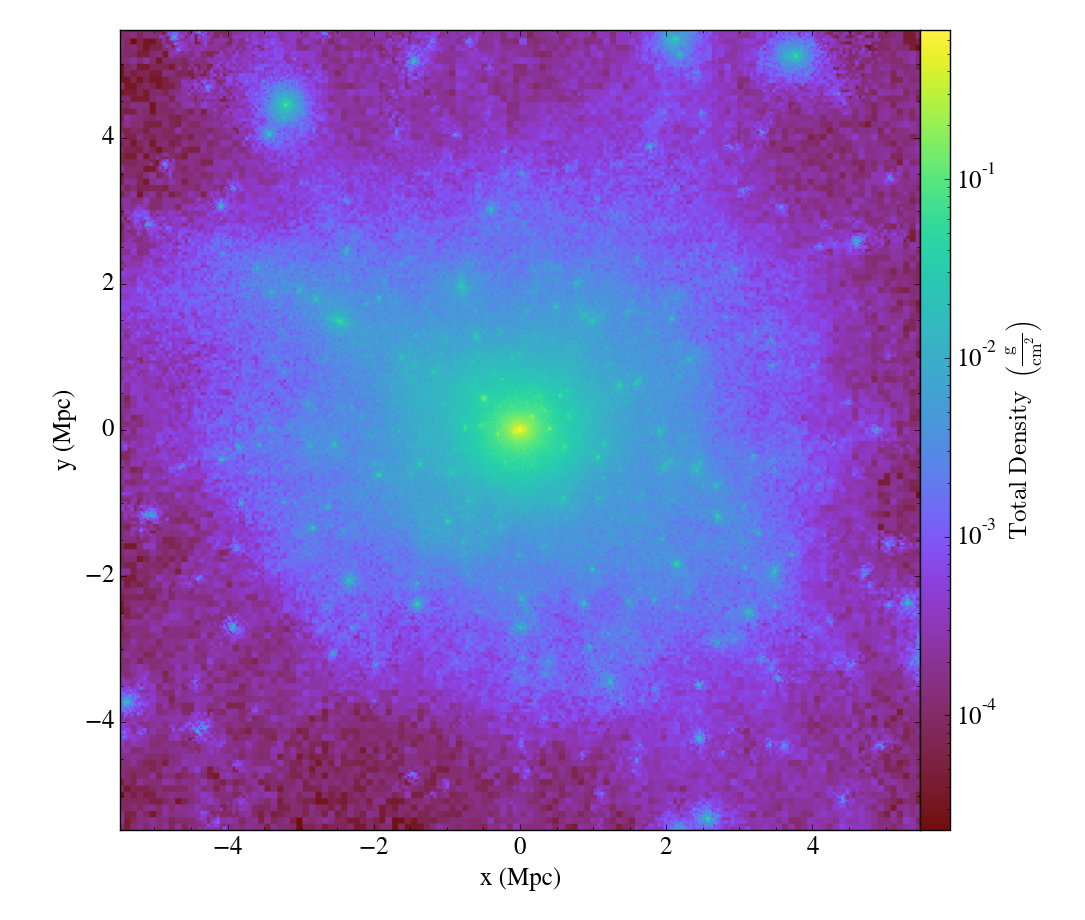}}
            \subfigure[]{\includegraphics[width=0.35\textwidth]{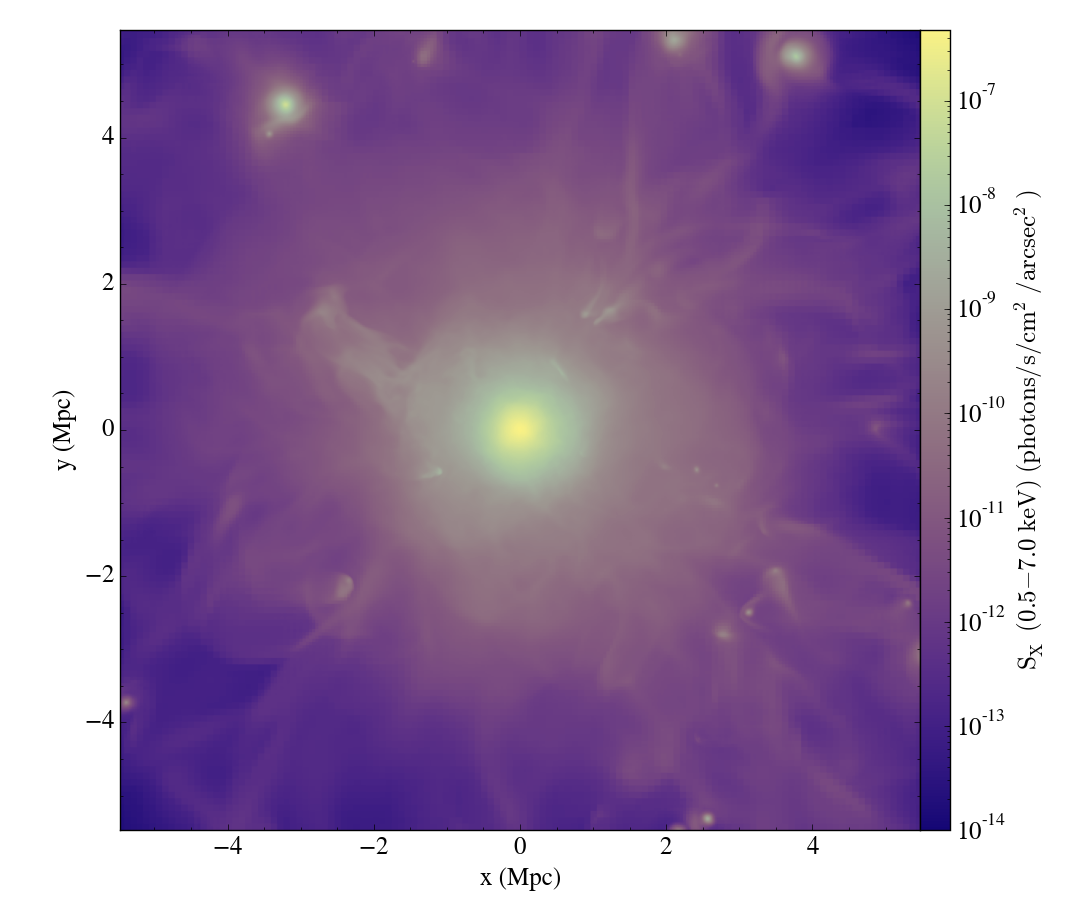}}        
            \subfigure[]{\includegraphics[width=0.3\textwidth]{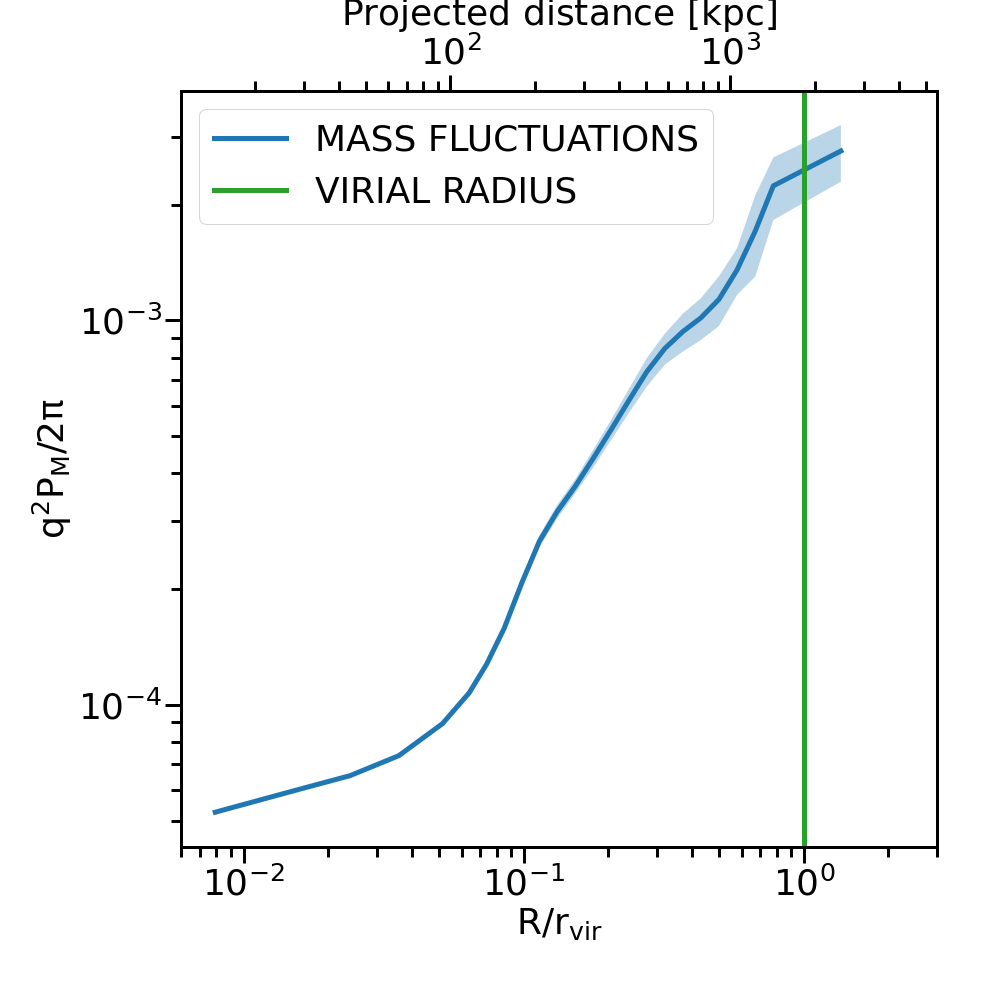}} 
            \subfigure[]{\includegraphics[width=0.3\textwidth]{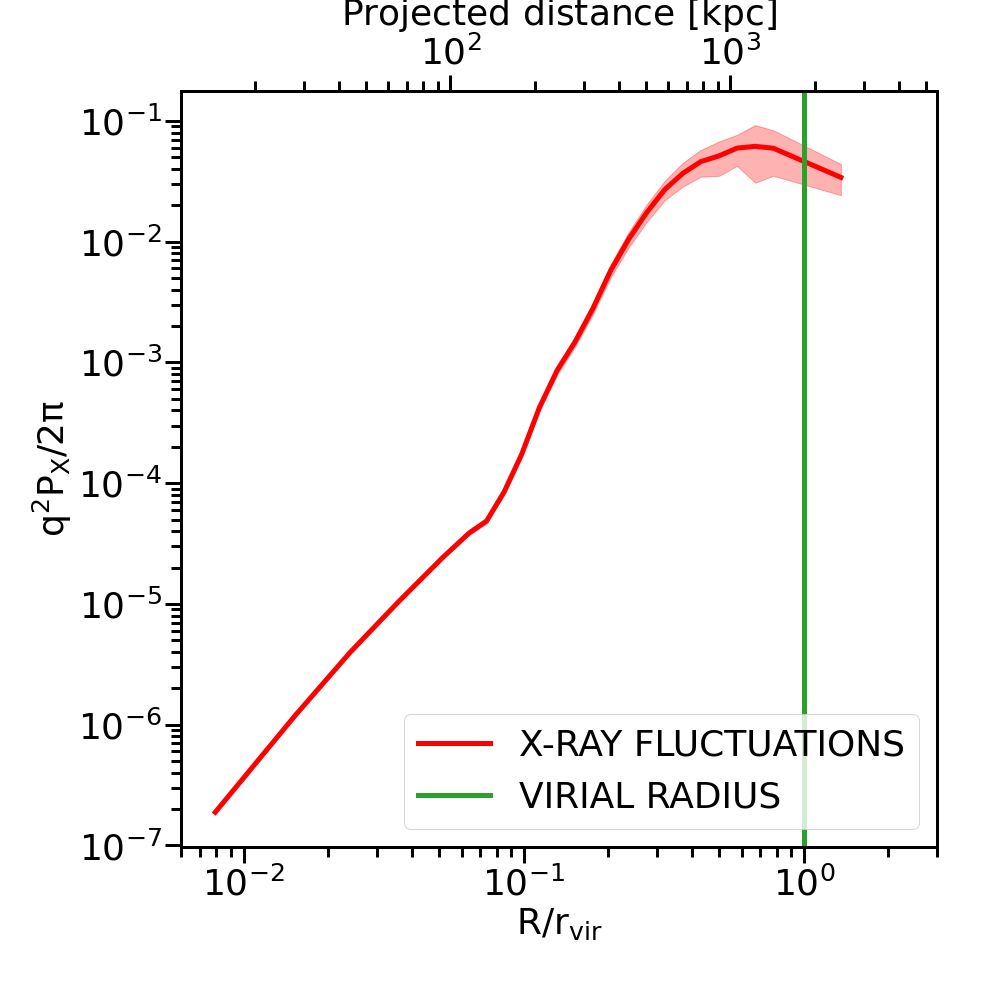}}
            \subfigure[]{\includegraphics[width=0.3\textwidth]{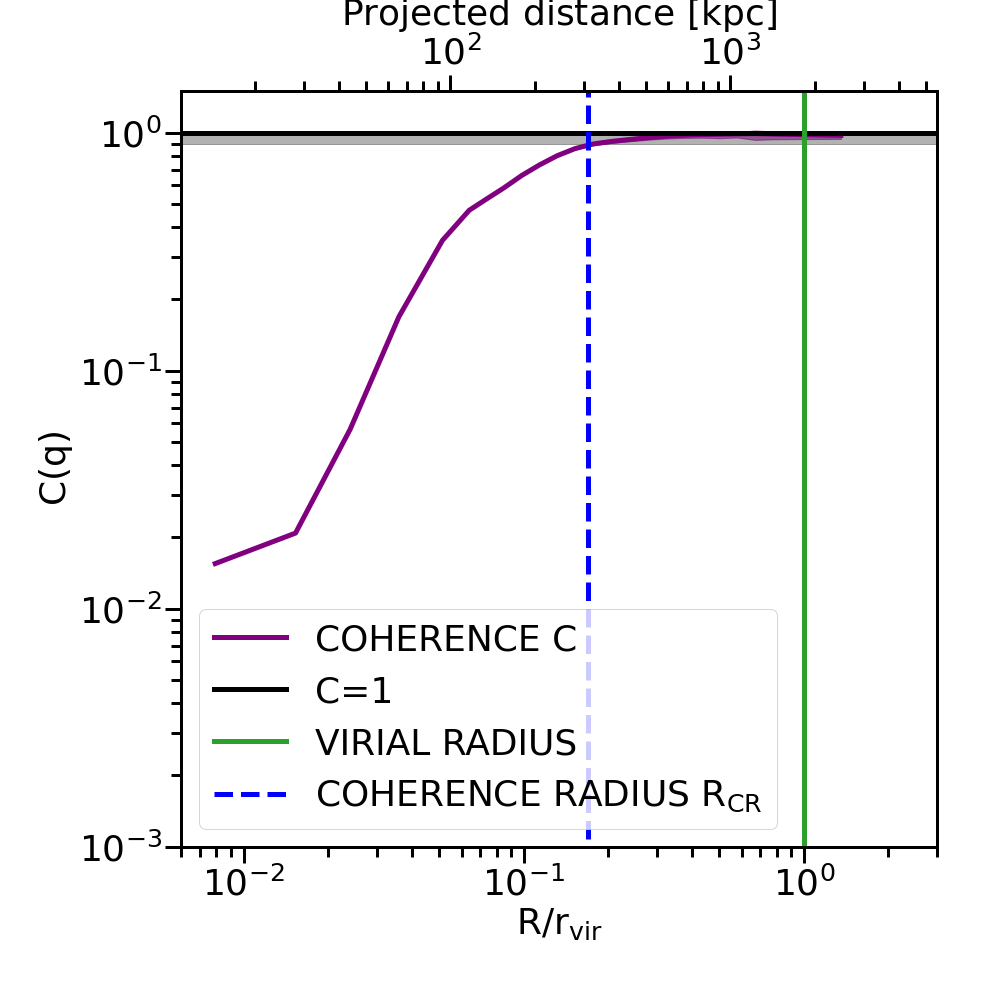}}
    \end{center}
    \caption{Analysis of the Simulated Cluster Analog C-3: (a) Mass map of the simulated cluster halo C-3; (b) X-ray surface brightness map of halo C-3; (c) computed mass dimensionless rms fluctuations of the halo C-3; (d) computed X-ray dimensionless rms fluctuations of halo C-3; (e) computed gas-mass coherence of halo C-3 as a function of scale. The bottom panels show the results as a function of the projected distance $R$ (top axis) and $R/r_{\rm vir}$ (bottom axis), with $r_{\rm vir}$ as virial radius.}
    \label{halo3}
\end{figure*}

The original images, shown in the top panels of Figs. \ref{halo1}, \ref{halo2}, \ref{halo3} and \ref{halo4}, have a FoV of $11.2\times 11.2\,{\rm Mpc}$ and a pixel size of $5.45\,{\rm kpc}$. For this investigation we wanted to exclude the areas outside the clusters to match the footprint of the observed clusters. Therefore, we created cut-out images with a FoV of $4.5\times 4.5\,{\rm Mpc}$ (as Abell 2744), centered on the cluster potential minimum and with the same pixel size. 

The current simulations only model gravitational physics and non-radiative hydrodynamics, while they do not take into account cooling and star formation, that are expected to have a non negligible effect only at very small scales, smaller than those investigated here.  

\section{The formalism: power spectrum analysis and gas-mass coherence} \label{formalism}

\subsection{Power spectrum and coherence} \label{ps_c}

From both the mass maps and the X-ray surface brightness maps, fluctuation fields relative to the background mean can be obtained:
\begin{eqnarray}
\delta F_m(x)=F_m(x)/<F_m>-1;
\end{eqnarray}
\begin{eqnarray}
\delta F_X(x)=F_X(x)/<F_X>-1;
\end{eqnarray}
\\
where $F_m(x)$ represents the mass values obtained from gravitational lensing for the observed clusters ( see Subsection \ref{lensing}) and from the OMEGA500 suite cluster halo mass maps from the simulations (see Subsection \ref{omega500}). Here, $F_X(x)$ are the masked, background-subtracted, exposure-corrected X-ray data for the observed clusters (see Section \ref{xray}) and the X-ray emissivity values coming from the OMEGA500 suite for the simulated clusters (see Section \ref{omega500}). $<F_m>$ and $<F_X>$ are the average values of the two datasets. For the observed  clusters Abell 2744 and Abell 383, the X-ray fluctuation fields were generated both for the signal $1/2(A+B)$ and for the noise $1/2(A-B)$ maps. In previous studies based on X-ray surface brightness fluctuations of galaxy clusters (see, for instance, \citealt{Churazov_2012} and \citealt{zhuravleva17}), the global cluster emission was removed by fitting a $\beta$-model to the images and dividing the images by the best-fitting models. This approach cannot be followed when the clusters are not in hydro-static equilibrium; an assumption that we want to interrogate with these new metrics.

Through the discrete FFT, provided by the python \textsl{numpy.fft} subpackage, the following Fourier transforms were then computed:

\begin{equation}
   \Delta_m (\textbf{q})=\int \delta F_m(x)\exp(-i\textbf{x}\cdot\textbf{q})d^2x.
\end{equation}
\\
\begin{equation}
   \Delta_X (\textbf{q})=\int \delta F_X(x)\exp(-i\textbf{x}\cdot\textbf{q})d^2x, 
\end{equation}

with $\textbf{x}$ coordinate vector in the real space, $\textbf{q}=2\pi \textbf{k}$ wave-vector, $|\textbf{k}|=1/\theta$, and $\theta$ angular scale. The 1D auto-power spectra can then be obtained:
 
 \begin{equation}
    P_m(q)=<|\Delta_m (\textbf{q})|^2>.
 \end{equation}
 \\
 \begin{equation}
    P_X(q)=<|\Delta_X (\textbf{q})|^2> 
 \end{equation}
 
where the average was taken over all the independent Fourier elements which lie inside the radial interval $[q,q+dq]$. From the Fourier transforms corresponding to the two different maps, we compute the cross-power spectrum:
 
\begin{IEEEeqnarray}{rCl}
P_{mX}(q)&=&<\Delta_m(q)\Delta_X^*(q)>\\
    \> &=&\textsl{Re}_m(q)\textsl{Re}_X(q)+\textsl{Im}_m(q)\textsl{Im}_X(q), \nonumber
\end{IEEEeqnarray} 

 with $\textsl{Re}$, $\textsl{Im}$ denoting the real and imaginary parts. The cross-power spectrum is a very powerful tool to measure the similarity of two signals as a function of space and has been widely adopted in the past decades in the analysis of multi-wavelength diffuse emission (see, for instance, \citealt{2012ApJ...753...63K}, \citealt{Cappelluti_2013}, \citealt{Helgason_2014}, \citealt{Li_2018} and \citealt{Kashlinsky_2018}). It is worth highlighting that the application of this tool to the study of galaxy clusters is not in conflict with the requirement of having stationary, but stochastic signals as suitable topics for the computation of the cross-power spectral density. Indeed, the evolutionary timescales of these astrophysical targets, as well as all the others investigated in the previous studies, range from millions to billions of years. Therefore, the emission under investigation here can be considered as stationary stochastic signals.
 
 The errors on the power spectra were obtained through the Poissonian estimators, defined as:
 
  \begin{equation}
    \sigma_{P_m}=P_m(q)/\sqrt{0.5N_q} 
 \end{equation}
 
  \begin{equation}
    \sigma_{P_X}=P_X(q)/\sqrt{0.5N_q} 
 \end{equation}
 
   \begin{equation}
     \sigma_{P_{mX}}=\sqrt{P_m(q)P_X(q)/N_q}
 \end{equation}
 
  with $N_q/2$ number of independent measurements of $\Delta_i (\textbf{q})$ out of a ring with $N_q$ data (since the flux is a real quantity and only one half of the Fourier plane is independent). For all clusters, the angular (and, consequently, the frequency) binning was optimized by choosing the bin sizes greater than the pixel sizes of the images and large enough to minimize the Poissonian error.
  
\begin{figure*}
    \begin{center}
            \subfigure[]{\includegraphics[width=0.35\textwidth]{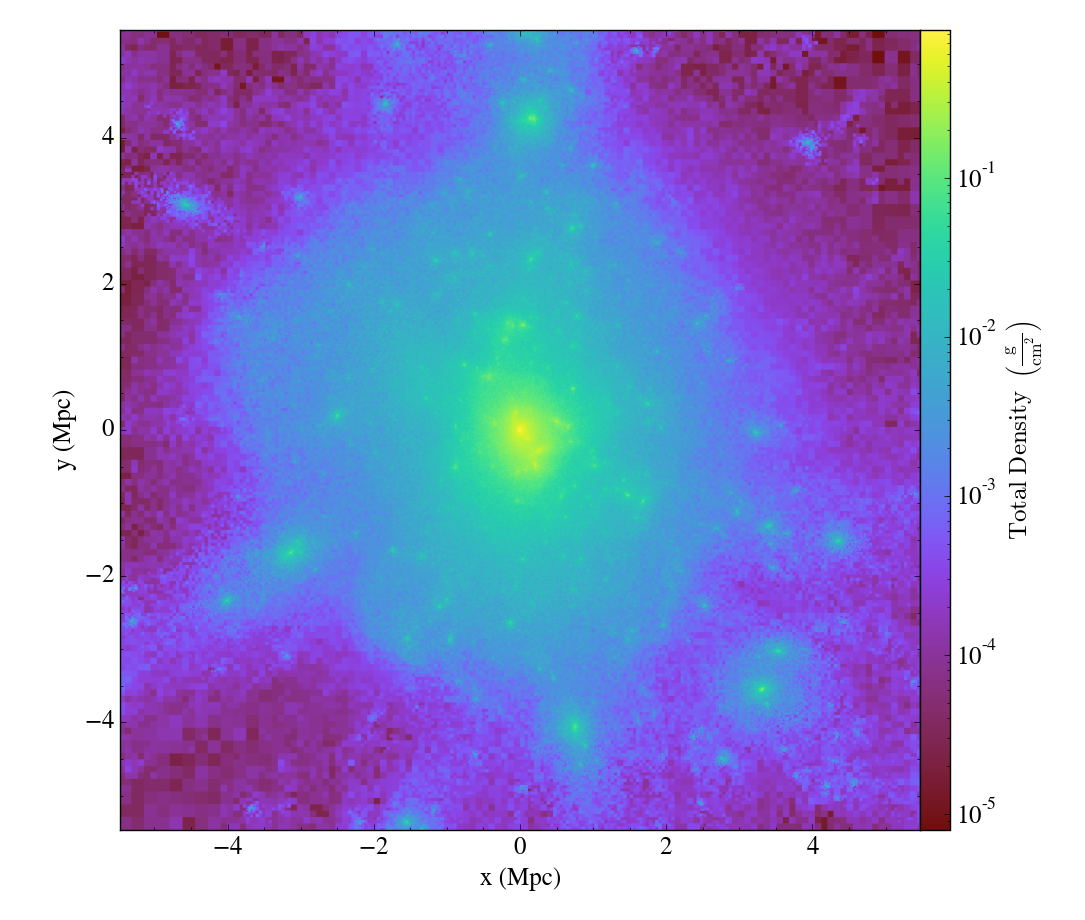}}
            \subfigure[]{\includegraphics[width=0.35\textwidth]{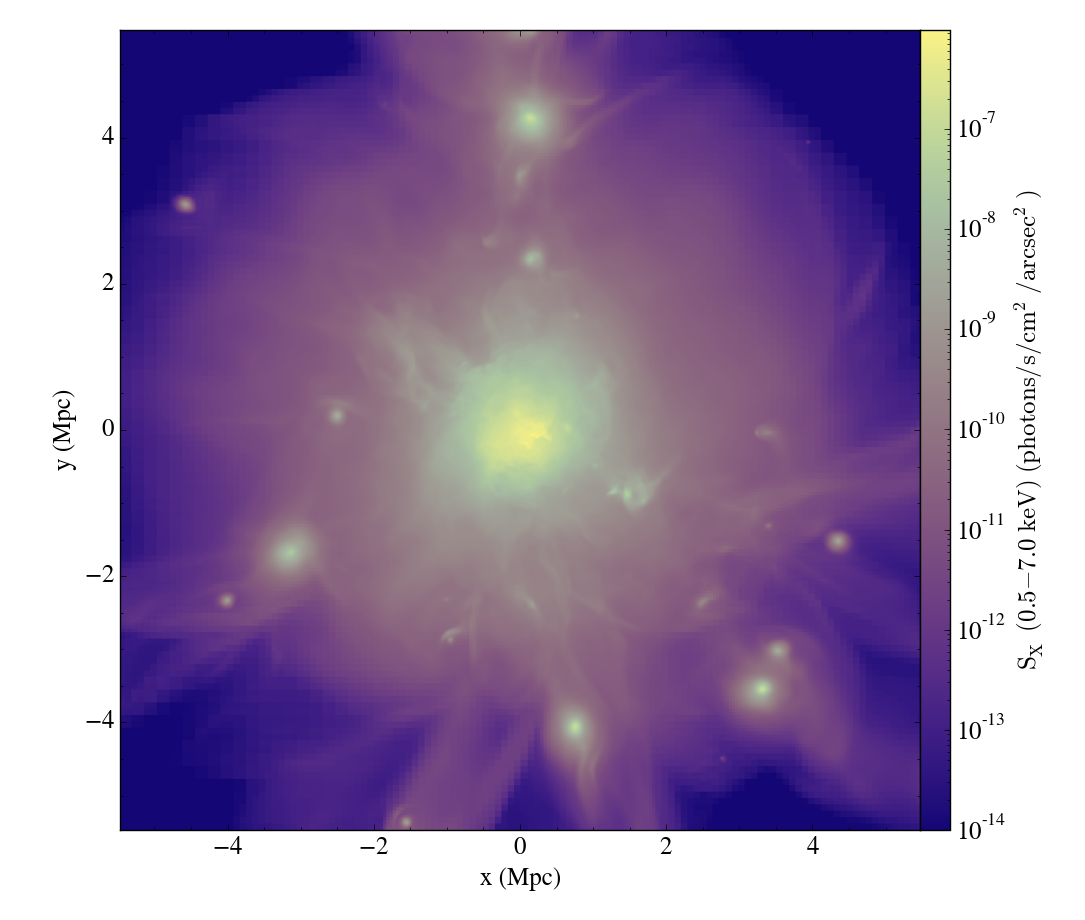}}        
            \subfigure[]{\includegraphics[width=0.3\textwidth]{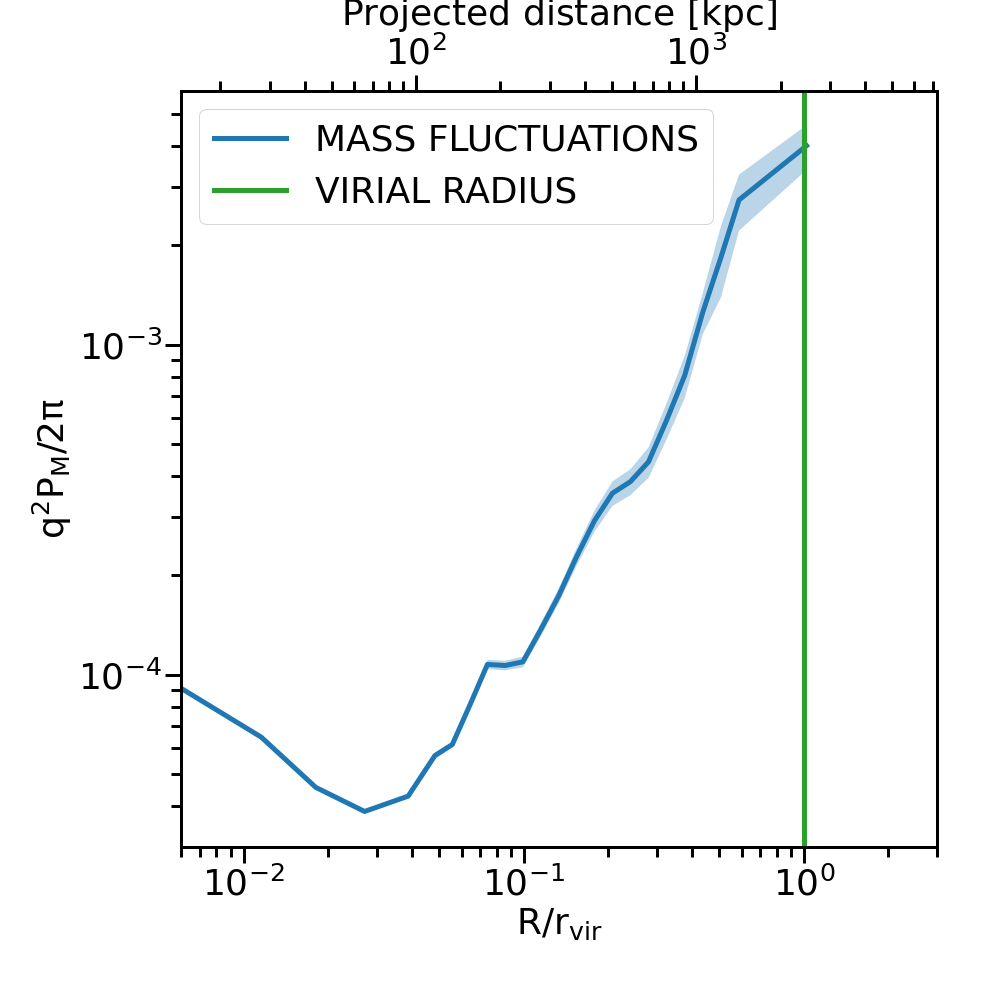}} 
            \subfigure[]{\includegraphics[width=0.3\textwidth]{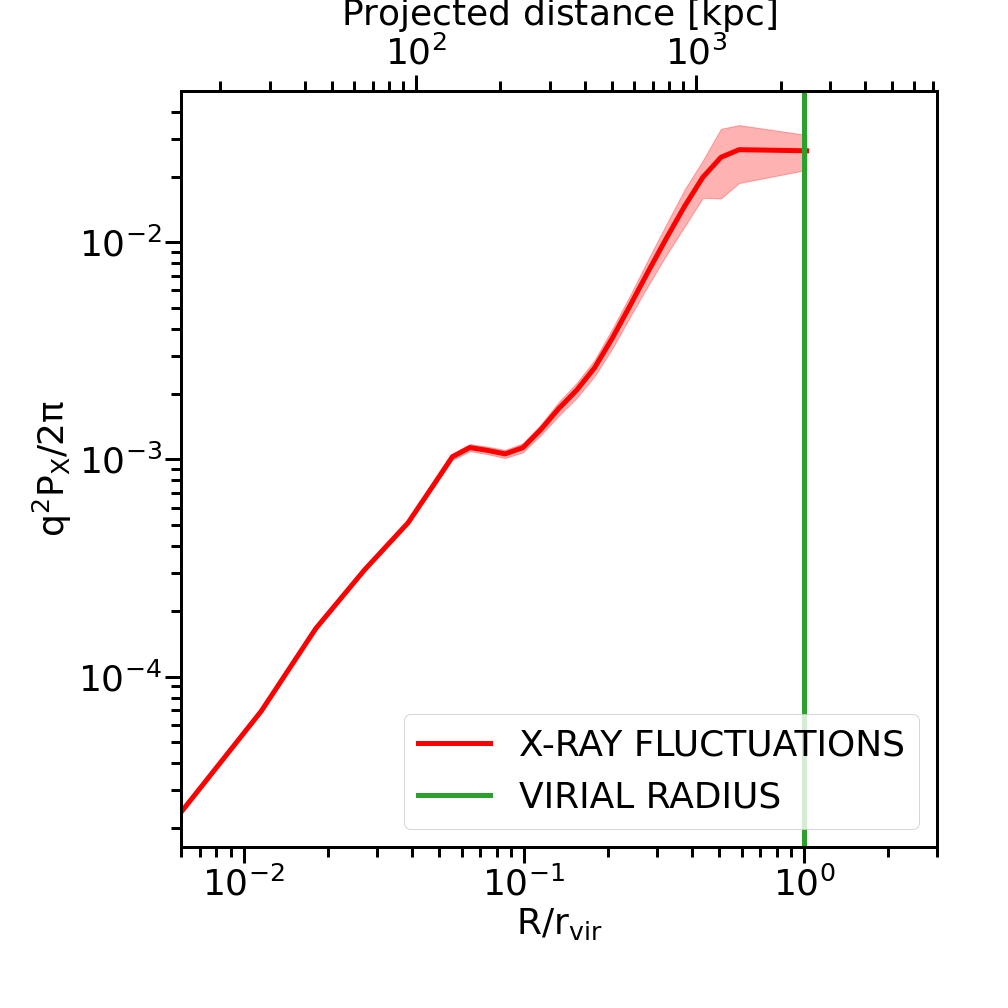}}
            \subfigure[]{\includegraphics[width=0.3\textwidth]{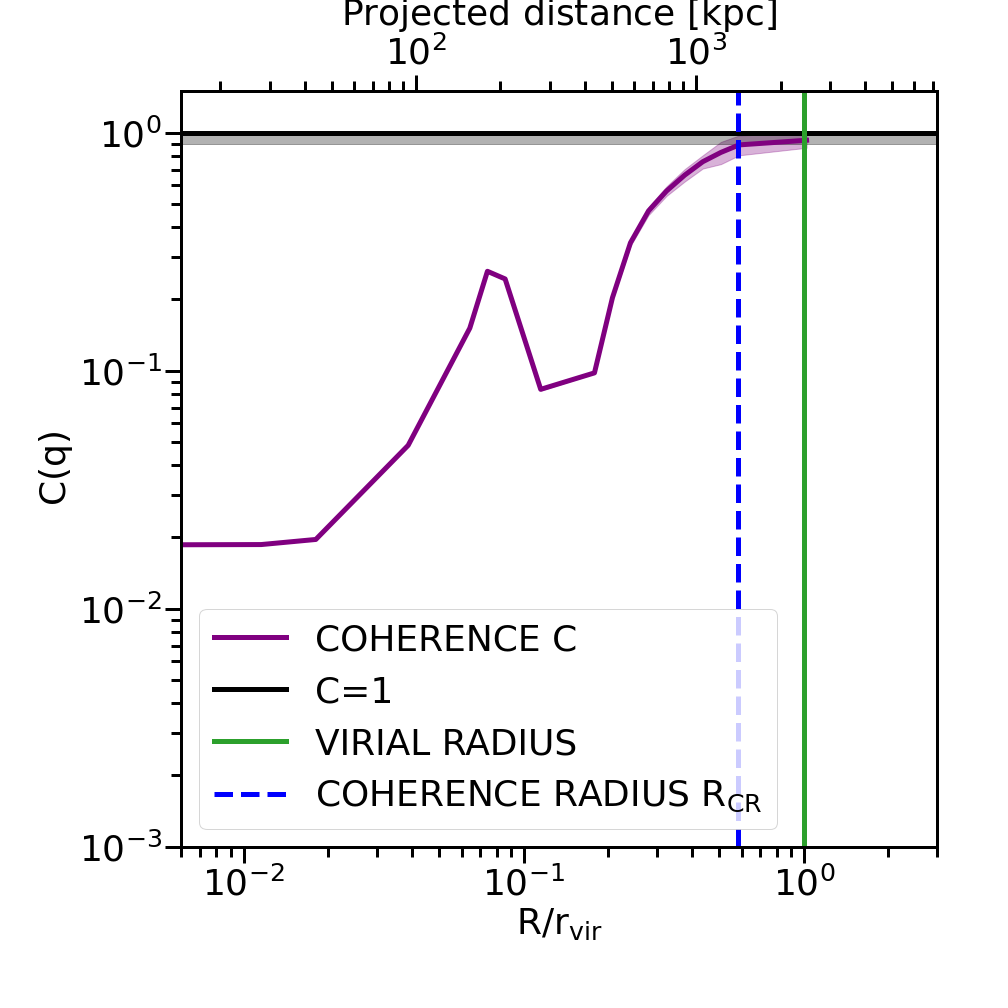}}
    \end{center}
    \caption{Analysis of the Simulated Cluster Analog C-4: (a) mass map of the simulated cluster halo C-4;(b) X-ray surface brightness map of halo C-4;(c) computed mass dimensionless rms fluctuations of halo C-4; (d) computed X-ray dimensionless rms fluctuations of halo C-4; (e) computed gas-mass coherence of halo C-4 as a function of scale. The bottom panels show the results as a function of the projected distance $R$ (top axis) and $R/r_{\rm vir}$ (bottom axis), with $r_{\rm vir}$ as virial radius.}
    \label{halo4}
\end{figure*}

  The rms fluctuations on scales $\theta=2\pi/q$ are usually computed as $\sqrt{q^2P(q)/2\pi}$, with $P(q)$ as the 1D power spectrum of the image under analysis. The simulated images have pixels in units of ${\rm kpc}$, thus the power spectrum is computed as a function of the projected distance $R$ (in units of ${\rm kpc}$). To be consistent with all the clusters in our study, for the observed cluster we converted the angular scales, in units of ${\rm arcsec}$, into a projected distance, in units of ${\rm kpc}$. As mentioned above, for Abell 2744, 1" corresponds corresponds to $4.54\,{\rm kpc}$, and for Abell 383, it corresponds to $3.15\,{\rm kpc}$. In addition, the power spectrum, computed with the \textsl{numpy.fft} subpackage from the dimensionless fluctuation fields, is dimensionless too. In order to have the dimensionless rms fluctuations, before computing $\sqrt{q^2P_m(q)/2\pi}$ and $\sqrt{q^2P_X(q)/2\pi}$, we rescaled $P_m(q)$ and $P_X(q)$ by $L^2/2\pi$, with $L^2$ FoV in ${\rm kpc^2}$. All the power spectra and rms fluctuations shown throughout the paper are dimensionless. The rms fluctuations and coherence obtained for our sample described in Table~\ref{table:1} are shown in Figs.~\ref{abell_2744}, \ref{abell_383}, \ref{halo1}, \ref{halo2}, \ref{halo3} and \ref{halo4} as a function of the projected radius $R$ (top axis) and the ratio $R/r_{\rm vir}$ (bottom axis), with $r_{\rm vir}$ as virial radius.
  
  Using the auto-power and cross-power spectra we can now compute a quantity that measures how well the gas properties reflect and trace the overall gravitational potential. This quantity is defined as the gas-mass coherence (or more generally, the coherence C), and was previously used in the multi-wavelength analysis of diffuse emission (see, for instance, \citealt{2012ApJ...753...63K}) and \citealt{Cappelluti_2013}:

   \begin{equation}
     C(q)=P_{mX}/\sqrt{P_m(q)P_X(q)}.
 \end{equation}
 
  A coherence equal to 1 at a specific scale corresponds to signals that are perfectly correlated or linearly related in structures at that scale, while a coherence equal to 0 indicates two totally uncorrelated signals or, as highlighted in \citealt{2012ApJ...753...63K} and \citealt{Cappelluti_2013}, if $C=1$, at the two channels the same populations produce the diffuse signal (or populations sharing the same environment).
  The gas-mass coherence, as we demonstrate in the following, is the key new tool we derive to assess the fidelity with which the gas traces the overall mass distribution. The coherence parameter offers an effective and efficient way to determine the level of correlation in Fourier space between fluctuations in the gas and the mass. We show that it is a powerful tool to spatially resolve features that are not obviously detectable through from images in real space and offers insights into how biased the gas is as a tracer of the potential. The coherence can be na\"{i}vely interpreted as a correlation coefficient between the spatial fluctuations in the two different maps. It is a clear-cut determinant of the validity of the hydro-static equilibrium. As highlighted more in details in Section~\ref{results}, in the event that the gas is relaxed and in hydro-static equilibrium with the underlying gravitational potential, the coherence will be unity.
 
 \subsection{Addressing Poisson noise and resultant errors} \label{poisson}
  
  As mentioned in Section \ref{xray}, the final power spectra $P_X$ of the the X-ray fluctuation fields for the real clusters were evaluated as $P_{1/2(A+B)}-P_{1/2(A-B)}$, with correspondingly propagated errors. Fig.~\ref{p_noise} shows the 1D dimensionless power spectra $P_{1/2(A+B)}$, $P_{1/2(A-B)}$ and $P_{1/2(A+B)}-P_{1/2(A-B)}$ in black, grey and red, respectively, for both Abell 2744 and Abell 383. The power spectrum of the difference image $1/2(A-B)$ is nothing other than the power spectrum of fluctuations due to Poisson noise. As expected, this noise spectrum has an essentially flat spectrum (white noise).
  
  \subsection{Addressing correction to the power spectrum from PSF effects} \label{psf}
  
 All the astronomical components of the power spectrum are affected by the instrument Point Spread Function (PSF) that acts like a window function. This translates into a multiplicative factor that applies to the effective power spectrum. This effect was modeled following the empirical approach proposed by \citealt{Churazov_2012}, thus multiplying at every frequency the true sky power spectrum by the following factor:
 
 \begin{equation}
    P_{PSF}(k)=\frac{1}{[1+(k/0.06)^2]^{1.1}}, 
 \end{equation}
  
 where $k$ is the angular frequency. As the PSF does not affect the particle background, the correction for the PSF blurring is applied by dividing the clean power spectrum $P_{1/2(A+B)}-P_{1/2(A-B)}$ by $P_{PSF}(k)$. 
 
 \subsection{Addressing shot noise due to faint point sources} \label{shot_noise}
 
\begin{figure}
  \centering
            \subfigure[]{\includegraphics[width=0.4\textwidth]{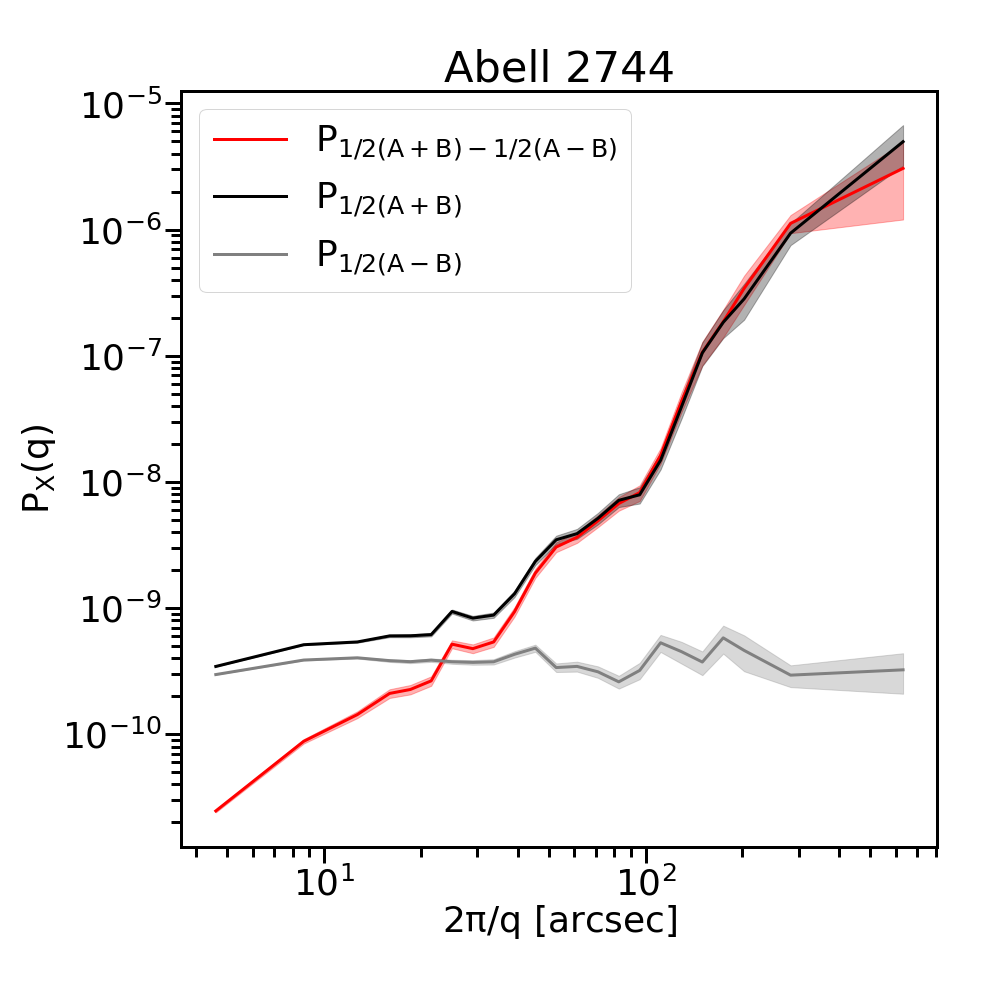}}
            \subfigure[]{\includegraphics[width=0.4\textwidth]{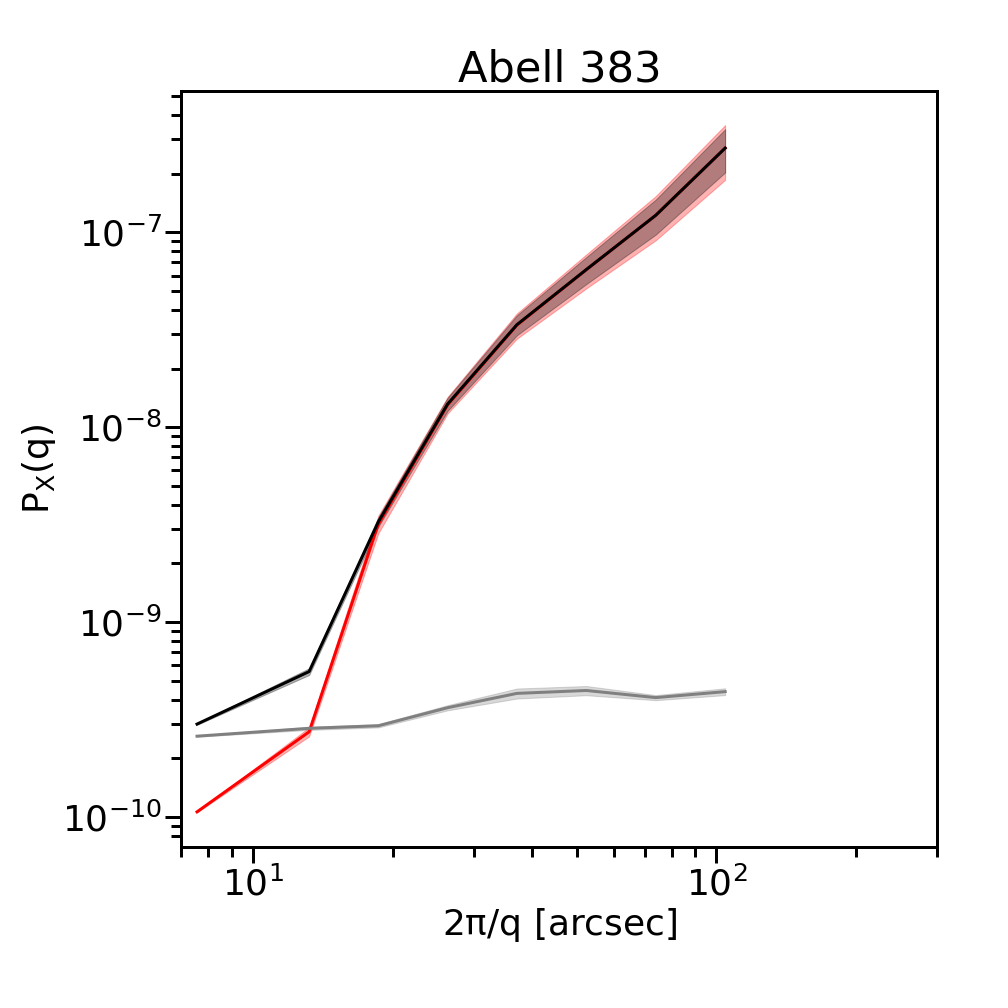}}   
            \caption{X-ray dimensionless auto-power spectrum results for Abell 2744 (upper panel) and Abell 383 (lower panel) as a function of the angular scale $\theta = 2\pi/q$. For both plots, the black curves show  $P_{1/2(A+B)}$, $P_{1/2(A-B)}$ is represented in grey, while the clean power power spectrum $P_{1/2(A+B)}-P_{1/2(A-B)}$ is displayed in red.}
 \label{p_noise}
 \end{figure}

 It is worth mentioning that, apart from bright detected sources, that were removed as explained in Section \ref{xray}, there do exist faint objects that are not detectable and yet add a shot noise component to the power spectrum. The contribution from these sources can be estimated starting from the Log N-Log S and limiting fluxes. In their study on X-ray brightness and gas density fluctuations in the Coma cluster, \citealt{Churazov_2012} showed that the Log N-Log S curves can be well approximated by the law $N \propto F^{-1}$, leading to the following contribution to the power spectrum:
 
 \begin{equation}
     P_{faint} \propto \int_{0}^{F_1} \frac{dN}{dF}F^2 \,df \propto F_1,
 \end{equation}
with $F_1$ lowest flux among the detected sources (the lowest counts we obtained are $\sim 30$ for Abell 2744 and $\sim 7$ for Abell 383). This term can be related to the power spectrum due to the bright sources, which in turn is given by:

 \begin{equation}
     P_{bright} \propto \int_{F_1}^{F_2} \frac{dN}{dF}F^2 \,df \propto F_2,
 \end{equation}
 if $F_2>>F_1$, where $F_2$ is the flux of the brightest detected source ($\sim 800$ counts for Abell 2744 and $\sim 100$ counts for Abell 383). Following the same approach, for both Abell 2744 and Abell 383, we obtained the contribution of unresolved sources at a level of the order of $\frac{F_1}{F_2}\sim 10^{-2}$ relative to the contribution of detected bright sources to the power spectrum, and therefore neglected this component. We want to highlight that the fact that the contribution from unresolved point sources can be a negligible component of the power spectrum is not surprising. Indeed, in many previous studies of the cosmic backgrounds (e.g. \citealt{Kashlinsky_2005}, \citealt{Kashlinsky_2012}, \citealt{Cappelluti_2012}, \citealt{Cappelluti_2013}, \citealt{Helgason_2014}, \citealt{Cappelluti_2017}, \citealt{Li_2018}, \citealt{Kashlinsky_2018}) it has been shown that there are two types of contributions relevant for the interpretation of the cosmological projected 2D power spectrum of source-subtracted fluctuations: the shot noise from remaining undetected sources that occasionally enter the beam, and the clustering component of the remaining cosmic background sources. The latter is generally composed of two terms (\citealt{CooraySheth_2002}): the 1-halo term, reflecting an average halo profile, and the 2-halo term, representing the halo-halo interaction. Nevertheless, our target is the diffuse emission coming from a single galaxy cluster and, therefore, one single halo, whose contribution to the power spectrum is reasonably expected to dominate the other terms. Finally, we need to comment on the scales at which the shot noise terms are relevant. Indeed, any shot noise contribution translates into a flat power spectrum component that, when non-negligible, visibly affects the power spectrum shape at small scales or, equivalently, large wavenumbers. In addition, it is affected by the telescope PSF and, consequently, it is flat at $k<<1/\sigma$ or, equivalently, $\theta>>\sigma$, and falls of at larger wavenumbers ($\sigma$ is the Gaussian width of the telescope PSF). From the shape of the red curves in Fig. \ref{p_noise}, representing the noise-subtracted power spectra, this feature of a flat component that falls of at small scales is not clearly visible. In addition, the fundamental tool of our investigation, as explained more in detail in Section \ref{results}, is the coherence analysis. Any eventual underestimation of the shot noise term due to faint point sources would lower the coherence at very small scales, and would not affect the main results of this investigation. 
   
 \subsection{Addressing edge effects in the data} \label{edge}
 
\begin{figure}
 \centering
            \subfigure[]{\includegraphics[width=0.4\textwidth]{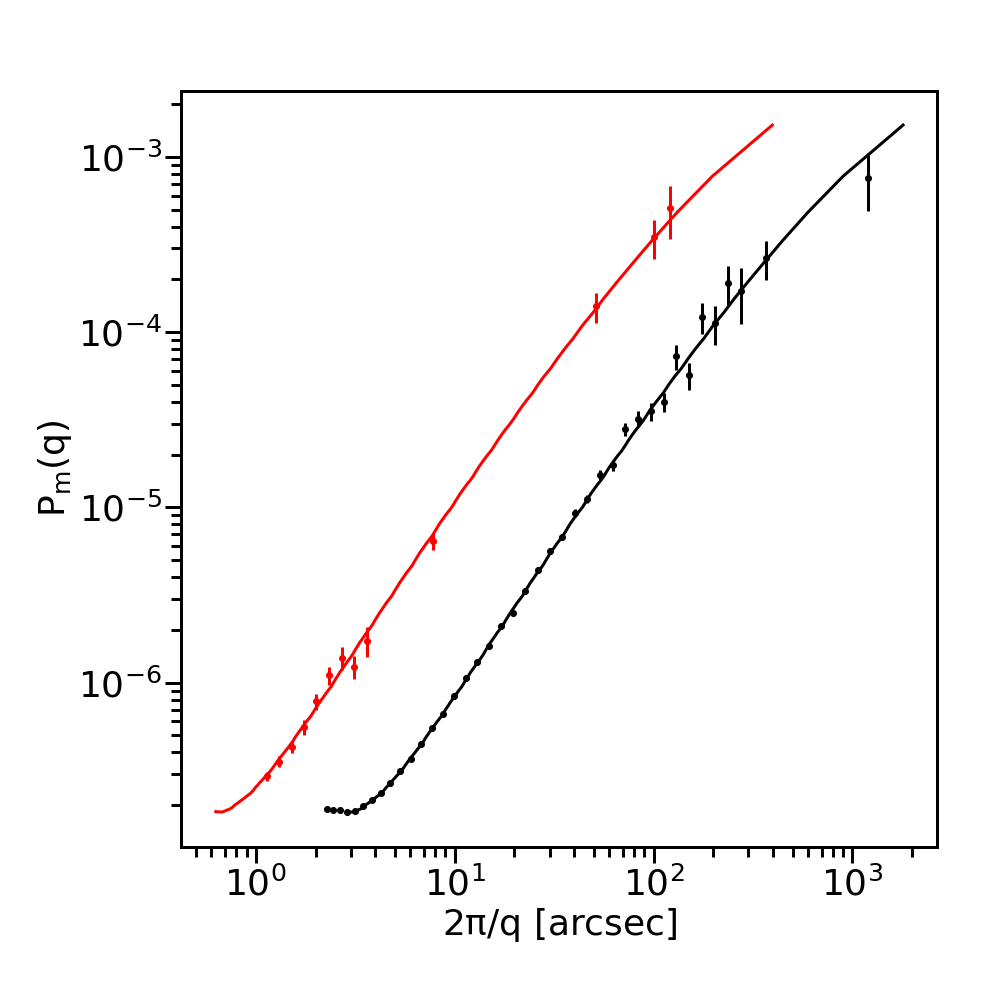}}
            \subfigure[]{\includegraphics[width=0.4\textwidth]{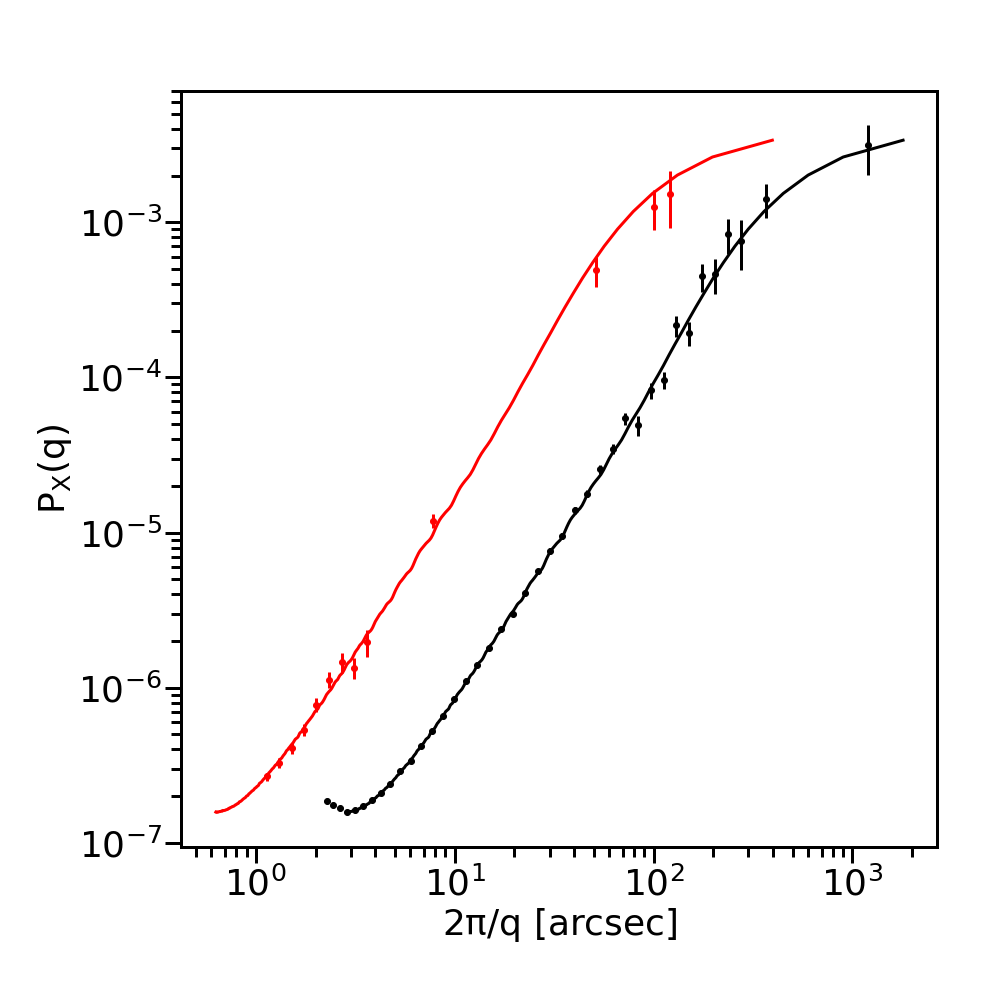}}   
            \caption{Dimensionless auto-power spectra obtained from NFW and $\beta$-profiles, compared with the auto-power spectra computed from the corresponding simulated images, as a function of the angular scale $\theta = 2\pi/q$. The curves obtained from the surface mass density are shown in the top panel, while the curves obtained from the X-ray surface brightness are shown in the bottom panel. The original auto-power spectra were computed from the NFW and $\beta$-profiles normalized to the central values, and are shown with continuous curves in both figures. The data with error bars represent the average auto-power spectra obtained from the simulated images. In red and black, we show the results obtained from the images with the same size of those created for Abell 383 and Abell 2744, respectively.}
 \label{edge_effect}
 \end{figure}

An additional question arises about the possibility that edge effects might affect the shape of the power spectra and, consequently, the coherence. To address this question, we use power spectra, computed from analytic NFW and $\beta$-profiles, to create images that were also subsequently analyzed through the developed Fourier transform and power spectrum analysis methodology presented above. In particular, we created two sets of NFW and $\beta$-profiles, choosing the virial masses in order to match those of our observed clusters Abell 2744 and Abell 383. From the power spectra of these two analytic models (examples of power spectra and coherence obtained from NFW and $\beta$-profiles are also shown in Section \ref{results}), we reconstructed two sets of images: 100 X-ray surface brightness and 100 surface mass density maps covering an area 5 times larger than those obtained from Abell 2744 and Abell 383 fields (100 X-ray maps and 100 mass maps per each cluster). From every large map, cut-out images of the same size of the real clusters under analysis were created and used to compute power spectra. Finally, for both clusters we computed the average power spectra, that were compared with the original power spectra obtained from the NFW and $\beta$-profiles. In Fig.~\ref{edge_effect} we show in red and black the results obtained from the images whose size matches that of Abell 383 and Abell 2744 fields, respectively. The continuous curves represent the original power spectra obtained from the analytic models, while the data with error bars refer to the average power spectra obtained from the simulated images. The top panel shows the results relative to the surface mass density, the bottom panel shows the results for the X-ray surface brightness. Since the data do not show any significant deviation from the curves, and the magnitude of any modification does not exceed the uncertainties in the power spectrum (dominated by Poisson noise), we can deduce that edge effects are negligible.

\section{Results} \label{results}

In Figs.~\ref{abell_2744} and \ref{abell_383} we show the results of the analysis with these new metrics for the observed clusters. In both figures the panels (a) and (b) on the top show the mass and X-ray maps respectively, while the panels (c), (d) and (e) on the bottom show the mass auto-power spectrum, the X-ray auto-power spectrum and the gas-mass coherence, respectively. Both the power spectrum analysis and the computation of the coherence reveal some important difference in the structure and evolutionary phase of these two clusters, in excellent agreement with the previous studies. Indeed, the results obtained for Abell 2744 reveal the highly disturbed dynamical state (previously investigated, for instance, by \citealt{Owers_2011}, \citealt{Merten_2011}, \citealt{10.1093/mnras/stw1435} and \citealt{Jauzac_2016}) and the presence of multiple substructures, while the data obtained for Abell 383 confirm the high level of relaxation and hydro-static equilibrium. 

The mass auto-power spectrum of Abell 2744 confirms the clumpiness of the gas and mass distribution, in agreement with the previous studies. In both the upper panels we show the scales of the main substructures, highlighted by the peaks in the power spectra displayed in the lower left and lower center panels of Fig.~\ref{abell_2744}, that are easy to distinguish in both maps. The distributions reveal the presence of significant substructures on scales of $120\,{\rm kpc}$, highlighted by the two small peaks on the same scales in the power spectra. A sub clump at the same scale is visible in the north-east corner of the X-ray image. In addition, the mass power spectrum also has a prominent feature on scales of $600\,{\rm kpc}$, in good agreement with the overall size of the Core feature. The X-ray power spectrum flattens on scales of about $1000\,{\rm kpc}$, that is about the size of the largest substructure in the gas distribution, while the mass power spectrum is much steeper at large scales. This difference in behaviour at large scales indicates that the mass is considerably more extended than the gas and the gas has collapsed in the central region. Another remarkable difference can be seen at very small scales, below $100\,{\rm kpc}$, the typical scale of individual galaxies. On these scales the X-ray power spectrum shows a prominent peak, whose main contribution is likely from the presence of the BCGs in the different sub clumps. The mass and X-ray power spectra in Fig.~\ref{abell_383}, for Abell 383, on the contrary, reveal a completely different picture. Most of all, the smoothness of the two curves is the clear signature that the two distributions - mass and gas - are not perturbed by the presence of substructures or disturbances, and this offers a clue to the fact that this system has not undergone a recent merger.

\begin{figure}
\includegraphics[width=1.0\linewidth]{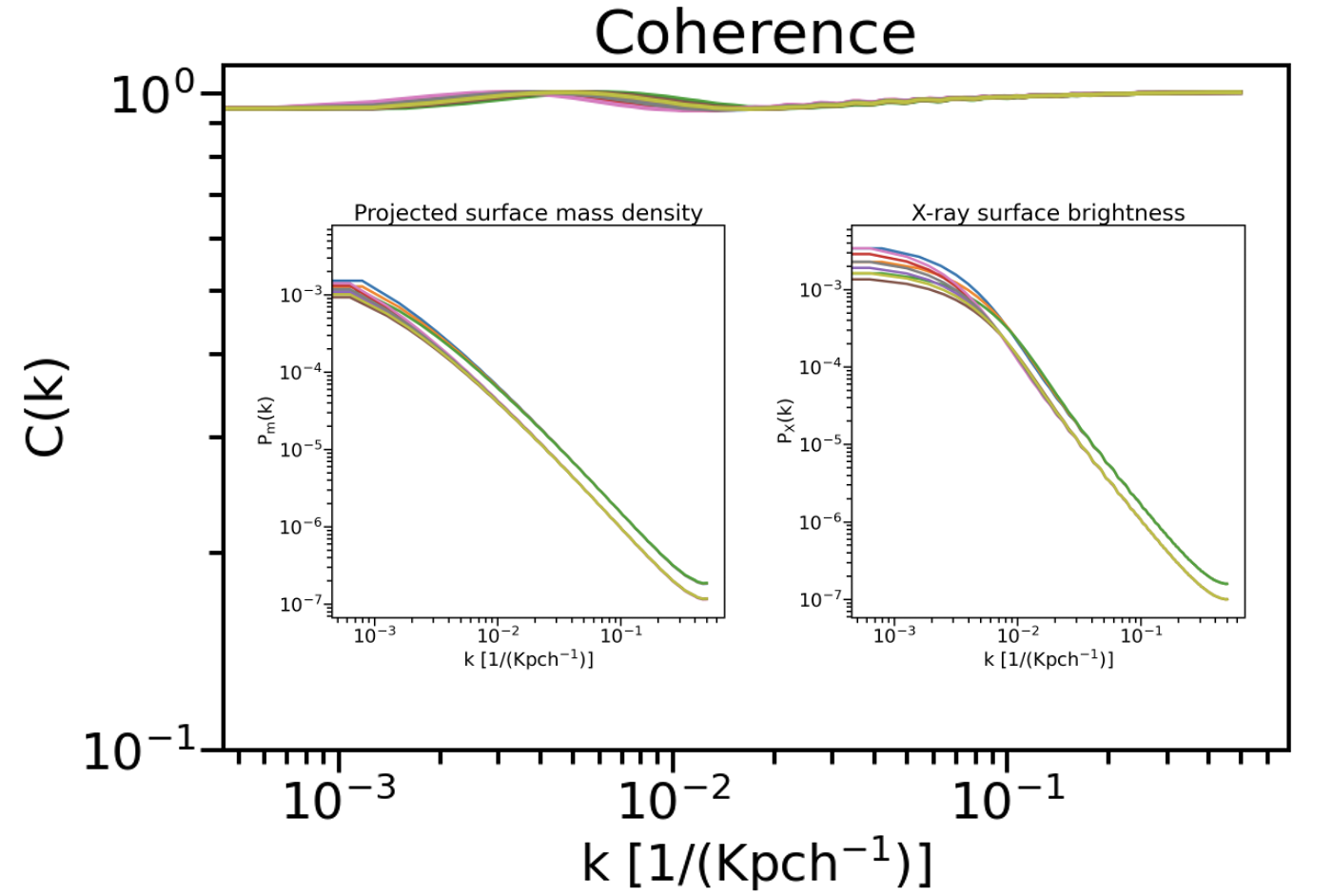}
\caption{Computed coherence for a set of virialized galaxy clusters with different values of virial mass $M_{\rm vir}$ and concentration $c$. The two smaller panels show the mass (on the left) and X-ray (on the right) dimensionless auto-power spectra, obtained from analytic NFW and $\beta$-profiles, all normalized to the central values. All the data are plotted as a function of the wave number $k=1/\theta$, with $\theta$ as angular scale.}
\label{theo_coh}
\end{figure}

As stated above, the key tool in our investigation is the gas-mass coherence. It has been recognized that, for virialized galaxy clusters and galaxy clusters in hydro-static equilibrium, the density profile can be well fitted by the NFW universal density profile (\citealt{Navarro_1996}, \citealt{Navarro_1997})

   \begin{equation}
     \rho(r)=\frac{\delta_c\rho_{c0}}{(r/r_s)(1+r/r_s)^2},
     \label{nfw}
 \end{equation}
 
 where $\rho_c=(3H^2(z))/(8\pi G)$ is the critical density of the Universe at redshift $z$ of the halo (with $H(z)$ Hubble space parameter at the same redshift and $G$ Newton constant), the scale radius $r_s=r_{200}/c$ is a characteristic radius at which the density profile agrees with the isothermal profile (i.e. $\rho(r) \propto r^{-2}$), $c=r_{\rm vir}/r_{\rm s}$ is the concentration and 
 
 \begin{equation}
     \delta_c=\frac{200}{3}\frac{c^3}{\ln(1+c)-c/(1+c)}
 \end{equation}
 
 is a characteristic over-density of the halo. From the volume density profile, we can derive the projected surface density:
 
 \begin{equation}
     \Sigma (R)=2 \int_{\infty}^{R} \frac{r\rho(r)}{\sqrt{r^2-R^2}} \,dr,
     \label{surf_d}
 \end{equation}
 
 with $R$ as the projected distance from the cluster center. The X-ray surface brightness profile, instead, is well approximated by a isothermal $\beta$-model (\citealt{1978A&A....70..677C})
 
 \begin{equation}
     I(R)=\frac{I_0}{[1+(\frac{R}{r_c})^2]^{3\beta -0.5}},
     \label{beta}
 \end{equation}
 where $I_0$ is the surface brightness at the center, $r_c$ is the gas core radius and $\beta$ is a fitting parameter (physically, the ratio of galaxy kinetic energy to gas energy). As explained in \citealt{Makino_1998}, the parameters in the equations \ref{nfw}, \ref{surf_d} and \ref{beta} are not independent and all the physical quantities characterizing the X-ray surface brightness and gas density profiles can be computed as a function of the total mass halo $M_{\rm vir}$ and concentration $c$. Computing the FFT and power spectra of the profiles in equations \ref{surf_d} and \ref{beta}, we derive the coherence, and find that it is basically a constant with a value of unity as function of scale (apart from small oscillations between 0.94 and 1). As an example, in Fig.~\ref{theo_coh} we show the computed coherence for a set of virialized galaxy clusters as a function of virial mass $M_{\rm vir}$ and concentration $c$. The power spectra obtained from the analytic NFW and $\beta$-profiles, all normalized to their central values, are plotted in the inner panels. The coherence for these relaxed clusters is close to unity at all scales.
 
The gas-mass coherence analysis offers an efficient and fast method to determine the level of correlation in Fourier space, because any non negligible deviation from unity immediately reveals the dynamical state of the cluster and the presence of out-of-equilibrium structures. We define the \textsl{Coherence Radius} $R_{\rm CR}$ the radius that reveals the scale above which the fluctuations of mass and X-ray surface brightness are 90$\%$ coherent or, equivalently, the scale above which their coherence ${\rm C}$ is 0.9. In Figs. \ref{abell_2744}, \ref{abell_383}, \ref{halo1}, \ref{halo2}, \ref{halo3} and \ref{halo4} $R_{\rm CR}$ is shown with the blue vertical dashed line (the region between 1.0 and 0.9 is indicated in gray). We chose to highlight the area corresponding to a coherence $0.9 < C < 1.0$ in order to be consistent with the average uncertainty of $10\%$ on the power spectra and coherence results. The \textsl{Coherence Radius}, compared with the virial radius of the cluster, gives us an immediate clue to the level of relaxation and equilibrium of the cluster. For instance, in Fig.~\ref{abell_383} the coherence computed for Abell 383 is very close to unity down to scales of $50\,{\rm kpc}$, corresponding to $\sim 0.02r_{\rm vir}$, confirming that this cluster can be considered relaxed (it is reasonable to assume that the coherence is still flat at larger scales, for which we do not have data). On the contrary, in Fig. \ref{abell_2744} the computed coherence for Abell 2744 reaches the value 0.9 only at scales very close to the virial radius, confirming that the cluster is highly unrelaxed. In addition, the lack of smoothness and the presence of peaks and depths reflects the clumpiness of the two distributions. For the set of simulated clusters we find a wider variety of cases. From the least to the most relaxed halo, we list the results below. For the halo C-2, as shown in the lower right panel of Fig. \ref{halo2}, the coherence $C$ never reaches the value $0.9$, showing that this galaxy cluster is totally unrelaxed. The presence of a merger is also highlighted by the presence of a secondary peak in the mass power spectrum shown in panel (c) on scales of $\sim 700\,{\rm kpc}$, the scales of the two substructures in the center of the image in panel (a). The X-ray power spectrum shown in panel (d), instead, does not exhibit the same peak but the curve flattens at smaller scales, between $\sim 300$ and $400\,{\rm kpc}$, highlighting that the X-ray emitting gas has collapsed in smaller structures with compared to dark matter. For the halo C-4 the value of the coherence is higher than 0.9 only at scales larger than $\sim 1400\,{\rm kpc}$, corresponding to $\sim 0.6r_{\rm vir}$. The X-ray power spectrum in panel (d) flattens at scales between $\sim 300$ and $400\,{\rm kpc}$, confirming the two central substructures highlighted by the contours in Fig.~\ref{contours}. The coherence computed for the halo C-1 is higher than $0.9$ down to scales of $\sim 530\,{\rm kpc}$, corresponding to $\sim 0.3r_{\rm vir}$, showing that the cluster is moving toward its relaxed phase. Finally, the halo C-3, the only simulated cluster that can be classified as relaxed using the X-ray contours, shows a coherence higher than $0.9$ down to scales of $\sim 310\,{\rm kpc}$, corresponding to $\sim 0.17r_{\rm vir}$. However, our coherence analysis reveals that the relaxed cluster C-3 has incoherent regions on scales whose size is $17\%$ of the virial radius, a non-negligible fraction of the entire cluster.

\begin{figure*}
    \begin{center}
        \includegraphics[width=0.9\textwidth]{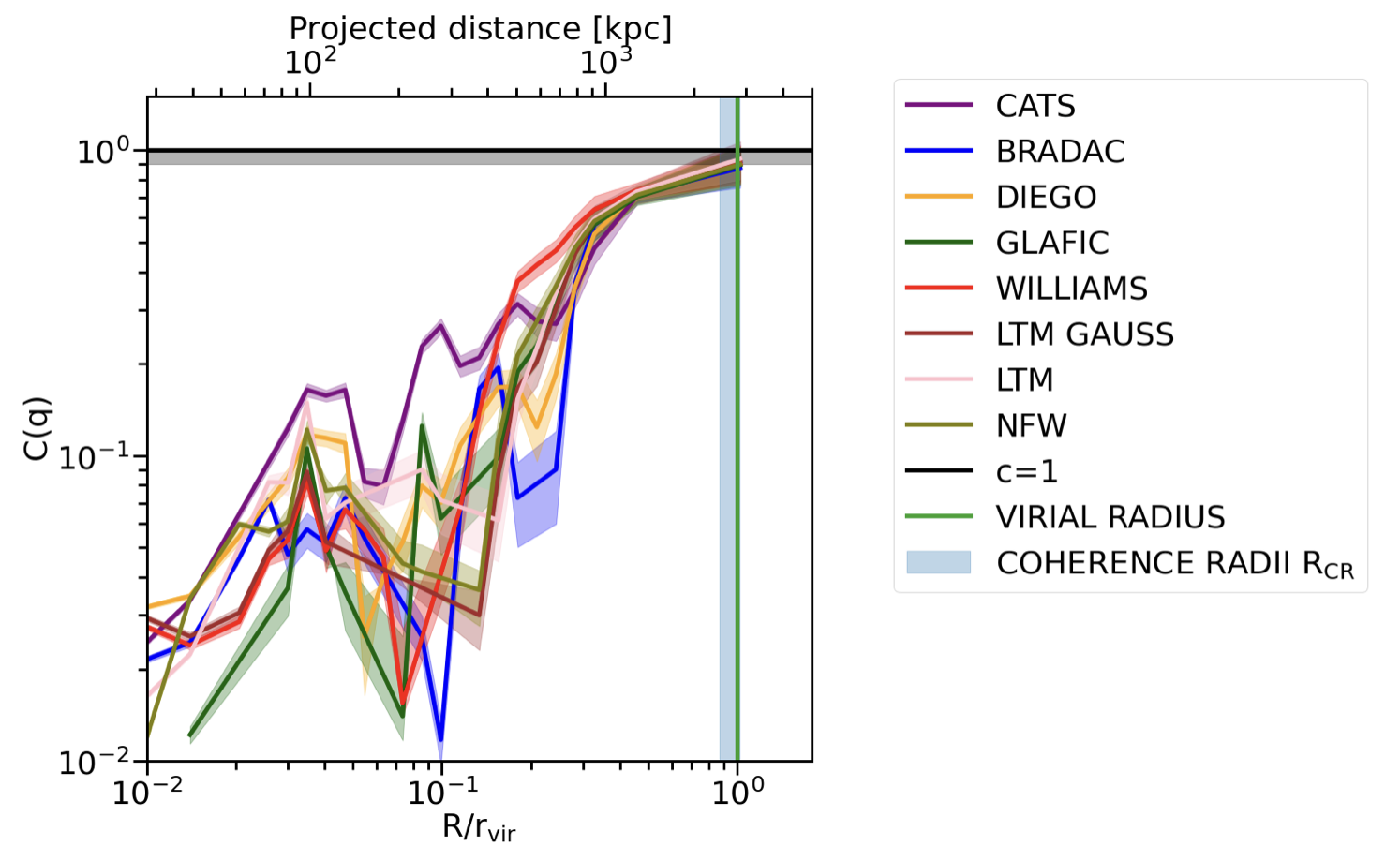}
    \end{center}
    \caption{Gas-mass coherence of Abell 2744, computed with different mass models: CATS (purple), BRADAC (blue), DIEGO (orange), GLAFIC (green), WILLIAMS (red), ZITRIN LTM GAUSS model (brown), ZITRIN LTM (pink) and ZITRIN NFW (brown). The vertical light blue shaded area highlights the range in which the \textit{Coherence radii} obtained from the different models are included.}
    \label{modelsa2744}
\end{figure*}

\begin{figure*}
    \begin{center}
        \includegraphics[width=0.9\textwidth]{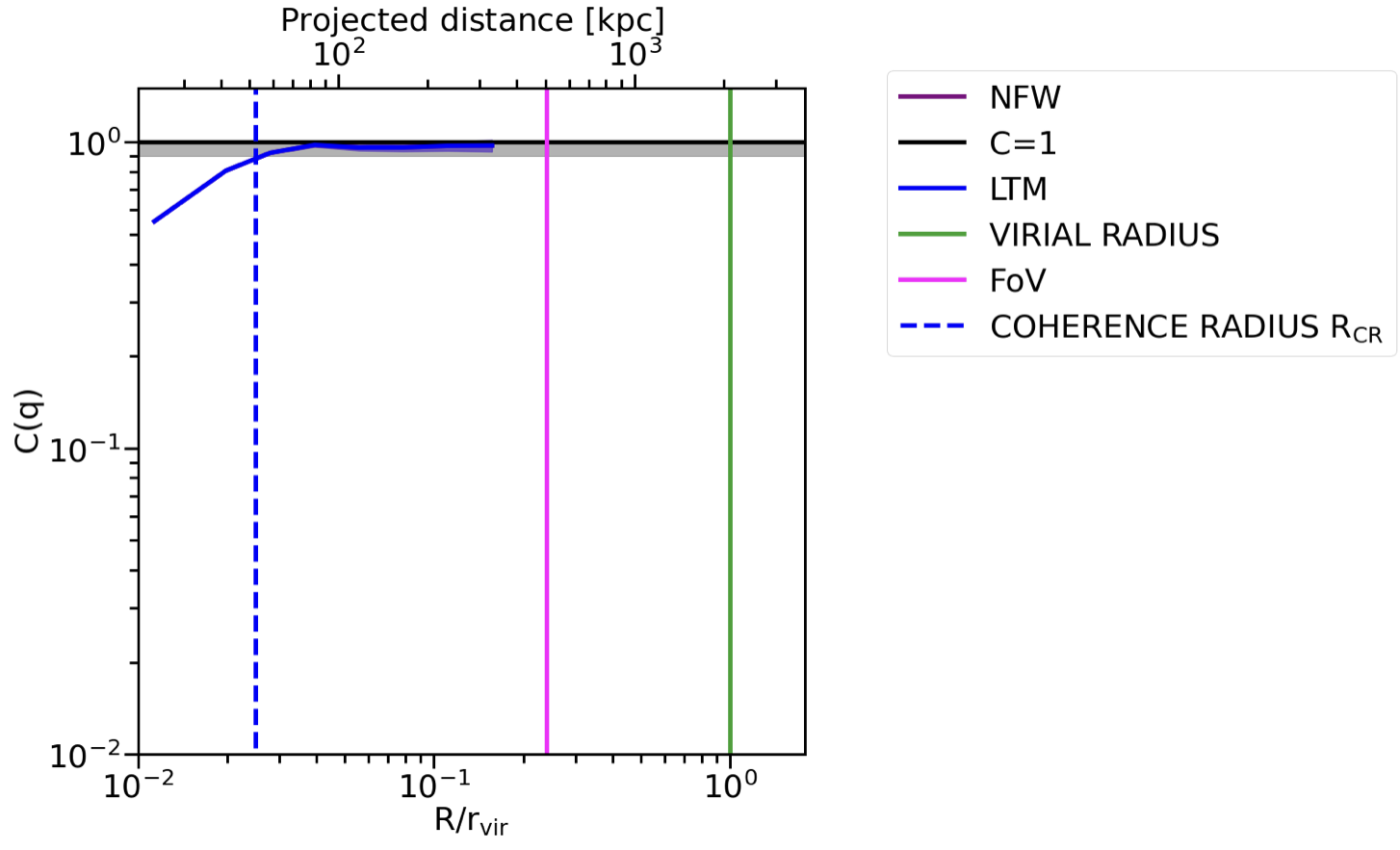}
    \end{center}
    \caption{Gas-mass coherence of Abell 383, computed with different mass models: ZITRIN NFW (purple) and ZITRIN LTM (blue). The vertival dashed line represents the \textit{Coherence Radius} (the \textit{Coherence Radii} computed for the two models coincide). }
    \label{modelsa383}
\end{figure*}

We finally investigate how the choice of different lensing mass maps, besides the CATS one, might affect our results. Figs. \ref{modelsa2744} and \ref{modelsa383} show the coherence computed for Abell 2744 and Abell 383, respectively, using the different mass reconstruction techniques described in Subsection \ref{lensing}. The differences in the power spectra, as well as the differences in the coherence analysis at scales below $R_{CR}$, require a deeper study and are being investigated in a separate paper. The fundamental result we want to highlight here is that, despite these differences, the key tool of our method, the \textit{Coherence Radius} $R_{CR}$, is not dramatically affected by the adopted mass model. In particular, the vertical light blue shaded area in Fig. \ref{modelsa2744} shows the range in which the \textit{Coherence Radii} obtained from the different models are included. This range corresponds to $\sim 15\%$ of the virial radius. In Fig. \ref{modelsa383} we plot only one dashed vertical line, since the \textit{Coherence Radii} overlap, and the coherence computed with the two different models does not show any significant difference.
 
\section{Discussion and Conclusions} \label{conclusions}

In this paper we present a new method for probing the dynamical state of galaxy clusters, based on the cross-correlation of the X-ray surface brightness and the mass distribution, the latter reconstructed from lensing observations. To demonstrate the proof of concept of this method and assess its reliability, we analyzed 2 observed clusters, the merging cluster Abell 2744 and the cool core relaxed cluster Abell 383, and 4 simulated clusters from the OMEGA500 suite. The X-ray maps of the observed clusters were cleaned of Poisson noise and spurious signals before being cross-correlated with the lensing data derived mass maps. The power spectrum and coherence analysis of fluctuations in these two maps led to the following main results.

\begin{itemize}
    \item The mass and X-ray auto-power spectra serve as new effective metrics that permit efficient and fast estimators to assess the level of clumpiness/smoothness in the mass and gas content of galaxy clusters, efficiently revealing the presence of substructures and on-going mergers, and their relevant physical scales. In particular, the results from the analysis of Abell 2744 and Abell 383 are in agreement with the results obtained from the prior studies: the former exhibits substructures both in the mass and in the gas distribution, the latter appears to have extremely smooth distributions, as expected from previous investigations.
    \item The new quantity gas-mass coherence that we have defined here can provide deeper insights into the dynamical state of galaxy clusters. The results obtained for Abell 2744 and Abell 383 confirm the unrelaxed status for the former and relaxed status for the latter.  The main new recipe to determine the degree of relaxation of galaxy clusters that we propose here is based on a quantity that we define as the \textsl{Coherence Radius} $R_{\rm CR}$ , that provides the scale at which fluctuations in the gas reflect those in the underlying mass to great fidelity, scales on which the coherence is therefore above 0.9. Using the results obtained from the analysis of simulated clusters from OMEGA500 cosmological suite, this radius appears to change according to the dynamical state and evolutionary phase of the cluster: the higher the level of disturbance, the more the $R_{\rm CR}$ moves towards larger scales. If the galaxy cluster is highly unrelaxed, the coherence never goes above 0.9. In the sample of simulated clusters analyzed here, the most disturbed cluster is the merging halo C-2, for which the coherence never rises above 0.9. For the other 3 halos - C-4, C-1 and C-3 - the $R_{\rm CR}$ gradually moves towards the smaller scales down to $0.6 r_{\rm vir}$, $0.3 r_{\rm vir}$ and $0.17 r_{\rm vir}$, respectively, highlighting that, even if a cluster appears to be in hydro-static equilibrium on larger scales, it still can have unrelaxed regions that are difficult to detect with other methods. It is worth highlighting that assessing the fidelity of the gas as a tracer of the underlying gravitational potential is deeply impactful and relevant for the use of clusters as cosmological probes, as all the cluster scaling relations that are used at present are predicated on the assumption of hydro-static equilibrium.
    \item Even though different mass reconstruction techniques that produce independent lensing maps can lead to differences in the coherence at scales below $R_{CR}$, they do not have a dramatic impact on the inferred value of the \textit{Coherence Radius} itself. This is a key result that shows the applicability of this new tool, because it reveals that the \textit{Coherence Radius} $R_{CR}$ is nearly model-independent. Take for example, the complex on-going merging cluster Abell 2744 -  the different mass models lead to an overall uncertainty on $R_{CR}$ of $\sim 15\%$, consistent with the other uncertainties involved in the investigation).
\end{itemize}

In the past, some relaxed galaxy clusters have been identified "by-eye", while other studies have proposed quantitative measurements of image features to assess their dynamical state: measures of bulk asymmmetry on intermediate scales and morphology in the X-ray imaging data (e.g. \citealt{1993ApJ...413..492M}, \citealt{Tsai_1996}, \citealt{Jeltema_2005}, \citealt{https://doi.org/10.48550/arxiv.1211.7040}, \citealt{2015MNRAS.449..199M}, \citealt{2016MNRAS.455.2936S}), identification of cool cores (e.g. \citealt{Vikhlinin_2007}, \citealt{Santos_2008}, \citealt{2009PhDT........18M}). The easy reproducibility and objectivity of these methods have led to a considerable improvement with respect to the visual classification. In particular, \citealt{2015MNRAS.449..199M} proposed an automatic method based on the symmetry, peakness and alignement of X-ray morphologies, that provides applicability to a wide range of redshifts and image depths. Nevertheless, the high resolution provided by the coherence analysis might reveal unrelaxed regions in clusters that were previously classified as relaxed (as in the case of the simulated cluster C-3). As explained at the end of this section, a follow-up paper with the results of this analysis extended to a wider sample of galaxy clusters is under preparation. Our results will be compared to those previously obtained from other methods.

In this work we have used the OMEGA500 simulations to widen our sample and extend the analysis to clusters in a larger variety of dynamical states. As we pointed out in sec. \ref{omega500}, these simulations do not take into account radiative hydrodynamics, that are expected to leave imprints in the power spectrum and, consequently, the coherence. Nevertheless, a detailed study of how all the features in the power spectrum and coherence reflect the different physical processes is beyond the scope of the current investigation but will be one of the topics we pursue in our extensive follow-up study, for which we will use other simulation suites, such as Magneticum or IllustrisTNG300. The method we introduce here is based on the new parameter $R_{CR}$, that quickly reveals the level of relaxation of the galaxy cluster. All the physical processes that cause departures from the equilibrium influence the power spectrum and coherence at scales smaller than the Coherence Radius $R_{CR}$.

Another question that we plan to address in follow-up papers is probing the role of projection effects. To investigate projection effects, we need to extend our framework to 2D and 3D power spectra. Nevertheless, we reiterate that, even with just the 1D power spectrum and cross-correlation analysis methods developed and presented here, any non negligible deviations of the coherence from unity offers a robust clue that the galaxy cluster is not in hydro-static equilibrium, as demonstrated with the analysis of the simulated cluster halos. Indeed, as highlighted in Section \ref{results}, the coherence of ideal relaxed clusters obtained from the projected mass and X-ray distributions is close to unity on all scales.

These new diagnostic metrics offer a wealth of possibilities for future follow-up investigations to extend the dynamic classification to galaxy clusters for which both lensing and X-ray maps are available. In particular, we are already working on the data reduction of the full HSTFF and CLASH samples, that offer targets with a wide range of masses and dynamical states out to redshift $z=0.89$. This wider range in redshift will also allow us to investigate if and how the depth of the X-ray observation influences the power spectrum measurements. For the X-ray analysis Chandra is still the best option for our investigation, since both XMM-Newton and eROSITA have a worse angular resolution. In addition, counting on a larger sample of observed clusters ranging in mass, dynamical states and at a wider range of redshifts, as well as additional simulation suites, we can study how the gas-mass coherence evolves during the cluster growth and assembly process. With this proof of concept demonstration of these new methods, it is clear that the prospects for more detailed and extensive applications to larger samples of clusters are promising.




\section*{Acknowledgements}
This research has been made use of data obtained from the Chandra Archives, HSTFF and CLASH mass maps, and simulated data from the Galaxy Cluster Merger Catalog (htpp://gcmc.hub.yt). The gravitational lensing models were produced by PIs Brada\v{c}, Natarajan \& Kneib (CATS), Merten \& Zitrin, Williams, Diego and the GLAFIC group. This lens modeling was partially funded by the HST Frontier Fields program conducted by STScI. STScI is operated by the Association of Universities for Research in Astronomy, Inc. under NASA contract NAS 5-26555.
The lens models were obtained from the Mikulski Archive for Space Telescopes (MAST). We gratefully acknowledge Mathilde Jauzac and Dominik Schleicher for providing us with the X-ray and lensing data for the Hubble Frontier Fields Cluster Abell 2744, as well as Erwin Lau for his support on our analysis on the simulated clusters from the OMEGA500 suite. GC and NC also acknowledge the University of Miami for the support. PN gratefully acknowledges support at the Black Hole Initiative (BHI) at Harvard as an external PI with grants from the Gordon and Betty Moore Foundation and the John Templeton Foundation.

\clearpage
\newpage

\bibliography{cluster_project}{}
\bibliographystyle{aasjournal}

\end{document}